\documentclass[canadian,english,nofootinbib, reprint]{revtex4-1}
\usepackage[T1]{fontenc}
\usepackage[latin9]{inputenc}
\usepackage{babel}
\usepackage{textcomp}
\usepackage{mathrsfs}
\usepackage{amsmath}
\usepackage{amssymb}
\usepackage{graphicx}
\usepackage[unicode=true]
 {hyperref}

\makeatletter

\providecommand{\tabularnewline}{\\}

\usepackage{braket}

\makeatother

\begin{document}
\selectlanguage{canadian}%
\global\long\def\d{\mathrm{d}}%

\global\long\def\braket#1#2{\Braket{#1|#2}}%

\global\long\def\bra#1{\Bra{#1}}%

\global\long\def\ket#1{\Ket{#1}}%

\global\long\def\kket#1{|#1\rangle\kern-0.25em  \rangle}%

\global\long\def\bbra#1{\langle\kern-0.25em  \langle#1|}%

\global\long\def\bbrakket#1#2{\langle\kern-0.25em  \langle#1|#2\rangle\kern-0.25em  \rangle}%

\global\long\def\alg{\text{UF}^{2}}%

\global\long\def\cvec#1{\mathbf{#1}}%

\renewcommand{\Im}{\text{Im}}

\providecommand{\Tr}{\mathrm{Tr}}
\title{Efficient numerical method for predicting nonlinear optical spectroscopies
\\of open systems}
\author{Peter A.~Rose }
\affiliation{Department of Physics, University of Ottawa, Ottawa, ON, K1N 6N5,
Canada}
\author{Jacob J.~Krich}
\affiliation{Department of Physics, University of Ottawa, Ottawa, ON, K1N 6N5,
Canada}
\affiliation{School of Electrical Engineering and Computer Science, University
of Ottawa, Ottawa, ON, K1N 6N5, Canada}
\begin{abstract}
Nonlinear optical spectroscopies are powerful tools for probing quantum
dynamics in molecular and nanoscale systems. While intuition about
ultrafast spectroscopies is often built by considering impulsive optical
pulses, actual experiments have finite-duration pulses, which can
be important for interpreting and predicting experimental results.
We present a new freely available open source method for spectroscopic
modeling, called Ultrafast Ultrafast ($\alg$) Spectroscopy, which
enables computationally efficient and convenient prediction of nonlinear
spectra, including treatment of arbitrary finite duration pulse shapes.
$\alg$ is a Fourier-based method that requires diagonalization of
the Liouvillian propagator of the system density matrix. We also present
a Runge-Kutta Euler (RKE) direct propagation method. We include open-systems
dynamics in the secular Redfield, full Redfield, and Lindblad formalisms
with Markovian baths. For non-Markovian systems, the degrees of freedom
corresponding to memory effects are brought into the system and treated
nonperturbatively. We analyze the computational complexity of the
algorithms and demonstrate numerically that, including the cost of
diagonalizing the propagator, UF\texttwosuperior{} is 20-200 times
faster than the direct propagation method for secular Redfield models
with arbitrary Hilbert space dimension; that it is similarly faster
for full Redfield models at least up to system dimensions where the
propagator requires more than 20 GB to store; and that for Lindblad
models it is faster up to dimension near 100, with speedups for small
systems by factors of over 500. $\alg$ and RKE are part of a larger
open source Ultrafast Software Suite, which includes tools for automatic
generation and calculation of Feynman diagrams.
\end{abstract}
\maketitle

\section{Introduction}

Nonlinear optical spectroscopies (NLOS) are widely used tools for
probing the excited state dynamics of a wide range of systems \cite{Abramavicius2009,Domcke2007}.
The signals that can be measured using NLOS contain a wealth of information,
but correctly interpreting that information generally requires making
a model of the system and predicting the spectra that result. Such
analysis can require repeated lengthy computations in order to fit
multiple parameters to the collected data \cite{Cho2005,Adolphs2007,Muh2007,Perdomo-Ortiz2012}.
Fast methods for simulating spectra of model system enable better
interpretation of experimental results. 

NLOS are often calculated in the impulsive limit of infinitely short
optical pulses. Recent work has shown that finite pulse effects can
have dramatic effects on measured NLOS, and that fitting experimental
data using intuition developed in the impulsive limit can lead to
incorrect conclusions \cite{Palecek2019}, adding to the existing
body of work exploring the effects of finite pulse shapes \cite{Faeder1999,Jonas2003,Belabas2004,Tekavec2010,Yuen-Zhou2012,Li2013,Cina2016,Do2017,Perlik2017,Smallwood2017,Anda2020,Suess2020}.
The effects of Gaussian and exponential pulse shapes have been treated
analytically for various types of NLOS, providing valuable insights
into the effects of pulse shapes and durations \cite{Smallwood2017,Perlik2017}.
However, real experimental pulses are often not well represented by
Gaussian or other analytical shapes. Ideally, modeling of NLOS should
include actual experimental pulse shapes rather than approximate forms,
and a number of numerical methods have this capability\textbf{ }\cite{Engel1991,Faeder1999,Belabas2004,Gelin2005,Gelin2009a,Renziehausen2009,Yuen-Zhou2014}.

In Ref.~\cite{Rose2019} we introduced a novel fast algorithm based
on Fourier convolution, called Ultrafast Ultrafast ($\alg$) spectroscopy,
capable of simulating any order NLOS using arbitrary pulse shapes.
We compared it to our own implementation of a standard direct propagation
method that we called RKE (Runge-Kutta-Euler) and demonstrated that
$\alg$ shows a significant speed advantage over RKE for systems with
a Hilbert space dimension smaller than $10^{4}$. However, that work
is based upon wavefunctions and is only valid for closed systems.
Condensed-phase systems consist of too many degrees of freedom to
treat them all explicitly, leading to essential dephasing and dissipation,
and making wavefunction methods of limited use in interpretation of
experiments.

In this work we present the extension of both $\alg$ and RKE to open
quantum systems with Markovian baths. Degrees of freedom corresponding
to memory in the bath can be included explicitly in the system Hamiltonian,
while the rest of the bath is assumed to be weakly coupled and treated
perturbatively using Redfield or Lindblad formalisms. We show that
$\alg$ is over 200 times faster than RKE for small system sizes,
and we believe this result is representative of the advantage that
$\alg$ provides over direct propagation methods. With a secular Redfield
model, $\alg$ outperforms RKE for all system sizes. Hereafter, the
terms $\alg$ and RKE refer to the new open extensions of the old
algorithms of the same name, with the understanding that the closed
system algorithms are now contained as special cases.

$\alg$ works in the eigenbasis of the Liouvillian that propagates
system density matrices and thus requires diagonalization of this
Liouvillian. We show that, surprisingly, the cost of this diagonalization
is negligible for the system sizes where $\alg$ outperforms RKE,
despite the Liouvillian having dimension $N^{2}$. Diagonalization
yields fast, exact propagation of the unperturbed system and allows
the optical pulses to be included using the computational efficiency
of the fast Fourier transform (FFT) and the convolution theorem. $\alg$
requires only that the pulse envelope be known at a discrete set of
time points, and thus is able to study any pulse shape of interest,
including experimentally measured pulse shapes. As few as 25 points
are required with Gaussian pulses to obtain 1\% convergence of spectra.

\begin{figure}
\includegraphics[width=\columnwidth]{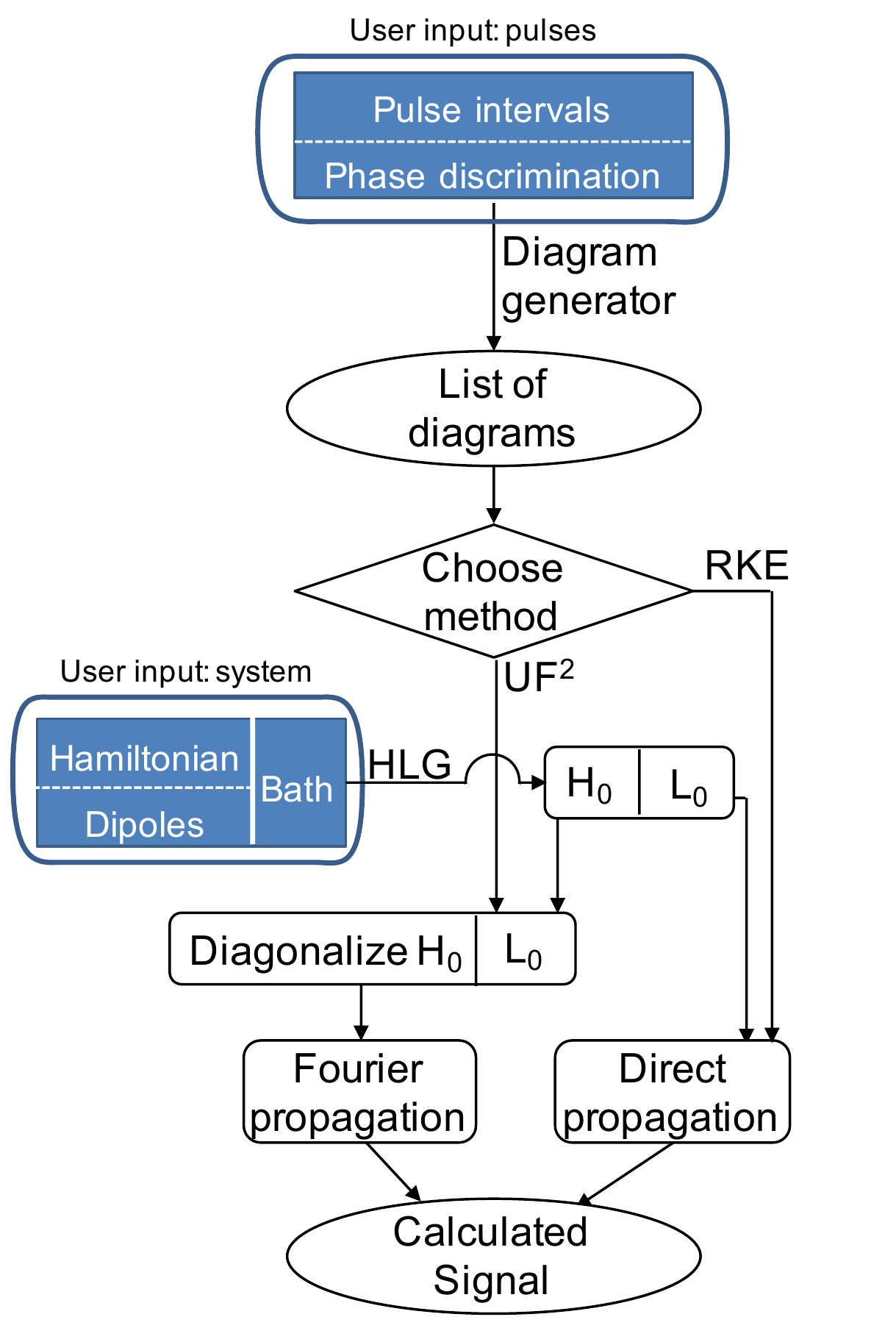}\caption{\label{fig:UFSS}Logical flow of the UFSS package. Users input information
about the experiment (pulse intervals and pulse-discrimination condition)
and the system (states, optical dipoles, and system-bath interaction).
UFSS consists of the diagram generator (DG), the Hamiltonian/Liouvillian
generator (HLG), and two choices of propagators: $\protect\alg$ and
RKE. The DG produces a list of Feynman diagrams, as described in Ref.~\cite{Rose2020a}.
This list and the Hamiltonian or Liouvillian of the system are inputs
to either $\protect\alg$ or RKE, which calculate the contribution
of each diagram to the resulting signal. The DG updates the list of
diagrams as the pulse delay times change, so that $\protect\alg$
and RKE only calculate causal diagrams for each set of pulse delays. }
\end{figure}

$\alg$ and RKE are part of a software package we call the Ultrafast
Spectroscopy Suite (UFSS), outlined in Fig.~\ref{fig:UFSS}, which
is designed to simplify the process of predicting spectra or fitting
spectra to models. UFSS is designed in particular to facilitate inclusion
of finite pulse effects with low computational cost. There are two
distinct effects of finite pulses. First is the inclusion of additional
Feynman diagrams that must be calculated when pulses overlap in time.
UFSS includes an automated Feynman diagram generator (DG), described
in Ref.~\onlinecite{Rose2020a}, which automates the construction
of these diagrams and determination of which ones give non-negligible
contributions. Second is the calculation of the contribution from
each diagram. Both $\alg$ and RKE take in diagrams and calculate
their contributions including the effects of pulse shapes. UFSS also
contains a Hamiltonian and Liouvillian generator (HLG), described
in this manuscript, which parametrically constructs models for vibronic
systems. Each of the packages in UFSS can be used independently. In
this work we demonstrate how $\alg$ and RKE can be used separately,
as well as with the HLG and DG. UFSS is free and open-source software
written in Python, available for download from \href{https://github.com/peterarose/ufss}{github}.

$\alg$ and RKE are numerical methods for including effects of optical
pulse shapes in the perturbative limit, given that the equations of
motion for an open quantum system in the absence of the pulses are
known. The efficient inclusion of finite pulse durations in $\alg$
relies on having a time-independent propagation superoperator for
the density matrix in the absence of optical fields. There are many
methods for describing the field-free dynamics of open quantum systems,
including Lindblad theory \cite{Gardiner2004}, Redfield theory \cite{Breuer2002},
multi configurational time-dependent Hartree (MCTDH) \cite{Raab1999,Raab2000},
and the hierarchical equations of motion (HEOM) \cite{Tanimura1989,Tanimura1990}.
Both Lindblad and Redfield theory allow treatment of dephasing and
relaxation due to a Markovian bath, resulting in time-independent
system propagators, and we implement both in $\alg$. While both HEOM
and MCTDH include non-Markovian bath dynamics, they do not yield time-independent
Liouvillians and are not amenable to the techniques in $\alg$; for
those methods, slower direct propagation methods are still required. 

In Sec.~\ref{sec:Algorithm} we briefly review the formalism of NLOS
calculated using time-dependent perturbation theory and then derive
the $\alg$ and RKE algorithms. The computational complexity of these
methods is shown in Appendix~\ref{sec:Computational-cost}. While
$\alg$ can propagate many types of systems, in Sec.~\ref{sec:Vibronic-Model}
we describe the HLG built-in to UFSS. In Sec.~\ref{sec:Computational-advantage}
compare the computational cost of $\alg$ and RKE for a range of system
sizes generated by the HLG. In Sec.~\ref{sec:Comparisons-to-literature}
we show the accuracy of $\alg$ by comparing to analytical expressions
for the 2D photon echo signal of the optical Bloch equations perturbed
by Gaussian pulses from Ref.~\onlinecite{Smallwood2017}. We demonstrate
that $\alg$ quantitatively agrees with the analytical results, including
effects of finite pulses, using just 25 evenly spaced points to represent
the Gaussian pulse shape.

\section{Algorithm\label{sec:Algorithm}}

We begin this section by outlining the standard results of time-dependent
perturbation theory, and how it is applied to nonlinear optical spectroscopies
\cite{Mukamel1999}, in order to introduce our notation and derive
the formal operators that we use to describe signals. In Sec.~\ref{subsec:UF2}
we build on this foundation to derive a novel open-systems algorithm
called $\alg$ for calculating perturbative spectroscopies. In Sec.~\ref{subsec:RKE}
we briefly present a direct propagation method called RKE that is
included in UFSS, which is used as a benchmark for timing comparisons
with $\alg$.

We begin with a Hamiltonian of the form
\begin{equation}
H=H_{0}(t)+H'(t),\label{eq:Htot}
\end{equation}
where the light-matter interaction with a classical field $\cvec E(t)$
is treated perturbatively in the electric-dipole approximation as
\begin{equation}
H'(t)=-\boldsymbol{\mu}\cdot\cvec E(t),\label{eq:Hprime}
\end{equation}
where $\boldsymbol{\mu}$ is the electric dipole operator. Cartesian
vectors are indicated in bold. We include a time-independent system-bath
interaction in the equations of motion for the system density matrix
$\rho$, so 
\begin{equation}
\frac{\d\rho}{\d t}(t)=-\frac{i}{\hbar}[H(t),\rho(t)]+D\rho(t),\label{eq:rhodot}
\end{equation}
where $D$ is a superoperator that describes dephasing and dissipation.
The $\alg$ algorithm can be applied with any time-independent operator
$D$. Separating the perturbation $H'(t)$ yields two superoperators,
$\mathscr{L}_{0}$ and $\mathscr{L}'(t)$, which are defined as
\begin{equation}
\frac{\d\rho}{\d t}(t)=\underbrace{\frac{-i}{\hbar}\left([H_{0},\rho(t)]+i\hbar D\rho(t)\right)}_{\mathscr{L}_{0}\kket{\rho(t)}}+\underbrace{\frac{-i}{\hbar}[H'(t),\rho(t)]}_{\mathscr{L}'(t)\kket{\rho(t)}}.\label{eq:rhodot1}
\end{equation}
$\rho$ can be considered as an operator in the Hilbert space of the
material system $\mathbb{H}$ and as a vector in the Liouville space
$\mathbb{L}$, which is the vector space of linear operators on $\mathbb{H}$.
We denote vectors in $\mathbb{L}$ by $\kket{\cdot}$. For linear
operators $A$ and $B$ acting on $\mathbb{H}$, we write the operator
$A\otimes B^{T}$ in $\mathbb{L}$ such that $A\otimes B^{T}\kket{\rho}$
is equivalent to $A\rho B$.\footnote{Note that the Liouville space is also a Hilbert space, with an inner
product $(v_{1},v_{2})$, which can be expressed in terms of the inner
product on $\mathbb{H}$. If $\kket{v_{1}}=\ket a\bra b$ and $\kket{v_{2}}=\ket c\bra d$,
then the inner product of $v_{1}$ and $v_{2}$ is $\Tr[v_{1}^{\dagger}v_{2}]=\braket db\braket ac$,
where the trace is taken with respect to a Hilbert-space basis. All
other cases following by linearity.} Using this transformation, we rewrite Eq.~\ref{eq:rhodot1} as

\begin{equation}
\frac{\d\kket{\rho(t)}}{\d t}=\mathscr{L}_{0}\kket{\rho(t)}+\mathscr{L}'(t)\kket{\rho(t)},\label{eq:EOM}
\end{equation}
where, in terms of operators on $\mathbb{H}$,
\begin{equation}
\mathscr{L}_{0}(t)=-\frac{i}{\hbar}H_{0}(t)\otimes\mathbf{\mathbb{I}}+\frac{i}{\hbar}\mathbb{I}\otimes H_{0}^{T}(t)+i\hbar D(t)\label{eq:L0}
\end{equation}
and
\begin{equation}
\mathscr{L}'(t)=-\frac{i}{\hbar}\boldsymbol{\mu}^{K}\cdot\mathbf{E}(t)+\frac{i}{\hbar}\boldsymbol{\mu}^{B}\cdot\mathbf{E}(t),\label{eq:Lprime}
\end{equation}
with
\[
\boldsymbol{\mu}^{K}=\boldsymbol{\mu}\otimes\mathbb{I}\quad\text{and}\quad\boldsymbol{\mu}^{B}=\mathbb{I}\otimes\boldsymbol{\mu}^{T}.
\]
In a closed system, $D=0$, and this formulation\textbf{ }becomes
equivalent to the closed case, which can be expressed with wavefunctions
rather than density matrices \cite{Rose2019}.

We describe the electric field as a sum over $L$ pulses, where each
pulse is denoted by a lowercase letter starting from $a$. A typical
$3^{rd}$-order signal is produced by up to 4 pulses. We write the
electric field as
\begin{equation}
\cvec E(t)=\sum_{j=a,b,...,L}\cvec e_{j}\varepsilon_{j}(t)+\cvec e_{j}^{*}\varepsilon_{j}^{*}(t)\label{eq:E(t)}
\end{equation}
where $\cvec e_{j}$ is the possibly complex polarization vector,
and the amplitude $\varepsilon_{j}$ of each pulse is defined with
envelope $A_{j}$, central frequency $\omega_{j}$, wavevector $\cvec k_{j}$,
and phase $\phi_{j}$ as
\[
\varepsilon_{j}(t)=A_{j}(t-t_{j})e^{-i\left(\omega_{j}(t-t_{j})-\cvec k_{j}\cdot\cvec r-\phi_{j}\right)},
\]
where $t_{j}$ is the arrival time of pulse $j$. We make the physical
assumption that each pulse is localized in time so $\varepsilon_{j}(t)$
is nonzero only for $t\in[t_{j,\text{min}},t_{j,\text{max}}]$. For
the purposes of UFSS, $A_{j}(t)$ does not need to be a closed-form
expression; it only needs to be known on a regularly spaced time grid
in $[t_{j,\text{min}},t_{j,\text{max}}]$. We define the Fourier transform
of the pulse as 
\[
\tilde{\varepsilon}_{i}(\omega)=\int_{-\infty}^{\infty}\varepsilon_{i}(t)e^{i\omega t}.
\]
The light-matter interaction, Eq.~\ref{eq:Lprime}, is a sum over
the rotating ($\varepsilon_{i}$) and counter-rotating ($\varepsilon_{i}^{*}$)
terms. We express these terms individually as 
\begin{align}
\mathscr{L}_{Kj^{(*)}}'(t) & =\frac{i}{\hbar}\boldsymbol{\mu}^{K}\cdot\cvec e_{j}^{(*)}\varepsilon_{j}^{(*)}(t)\label{eq:HKj}\\
\mathscr{L}_{Bj^{(*)}}'(t) & =-\frac{i}{\hbar}\boldsymbol{\mu}^{B}\cdot\cvec e_{j}^{(*)}\varepsilon_{j}^{(*)}(t)\label{eq:HBj}
\end{align}
so that
\begin{equation}
\mathscr{L}'(t)=\sum_{i=a,b,...}\mathscr{L}'_{K\,i}(t)+\mathscr{L}'_{K\,i^{*}}(t)+\mathscr{L}'_{B\,i}(t)+\mathscr{L}'_{B\,i^{*}}(t).\label{eq:Lp-decomposition}
\end{equation}
In the rotating wave approximation (RWA), the rotating terms, $\mathscr{L}'_{K\,i}$
and $\mathscr{L}'_{B\,i}$, excite the ket-side and de-excite the
bra-side of the density matrix, respectively. The counter-rotating
terms, $\mathscr{L}'_{K\,i^{*}}$ and $\mathscr{L}'_{B\,i^{*}}$,
excite the bra-side and de-excite the ket side, respectively \cite{Mukamel1999,Yuen-Zhou2014}.

We treat the effect of $\mathscr{L}'(t)$ using standard time-dependent
perturbation theory and assume that at time $t_{0}$ the system is
in a stationary state of $\mathscr{L}_{0}$, which is $\kket{\rho^{(0)}}$.
Equation \ref{eq:EOM} is easily integrated in the absence of perturbation
to give the time-evolution due to $\mathscr{L}_{0}$, 
\begin{equation}
\mathcal{T}_{0}(t)=\exp\left[\mathscr{L}_{0}t\right].\label{eq:Tdef}
\end{equation}
The perturbation $\mathscr{L}'(t)$ is zero before $t_{0}$ and produces
a time-dependent density matrix $\kket{\rho(t)}$, which is expanded
perturbatively as 
\begin{equation}
\kket{\rho(t)}=\kket{\rho^{(0)}}+\kket{\rho^{(1)}(t)}+\kket{\rho^{(2)}}+...\label{eq:rho-expansion-1}
\end{equation}
where the $n^{\text{th}}$ term can be expressed as \cite{Mukamel1999}
\[
\kket{\rho^{(n)}(t)}=\mathcal{T}_{0}(t)\int_{0}^{\infty}dt'\mathcal{T}_{0}^{-1}(t-t')\mathscr{L}'(t-t')\kket{\rho^{(n-1)}(t-t')}.
\]
Using the decomposition of $\mathscr{L}'(t)$ in Eq.~\ref{eq:Lp-decomposition},
we write $\kket{\rho^{(n+1)}(t)}$ as a sum over four types of terms
\begin{align*}
\kket{\rho^{(n+1)}(t)} & =\sum_{j}\left(K_{j}+K_{j^{*}}+B_{j}+B_{j^{*}}\right)\kket{\rho^{(n)}(t)},
\end{align*}
where all four terms are compactly defined as
\begin{equation}
O_{j^{(*)}}=\eta_{O}\frac{i}{\hbar}\mathcal{T}(t)\int_{0}^{\infty}dt'\mathcal{T}^{-1}(t-t')\left(\boldsymbol{\mu}^{O}\cdot\cvec e_{j}^{(*)}\varepsilon_{j}^{(*)}(t-t')\right),\label{eq:Odef}
\end{equation}
with $O=K,B$, $\eta_{K}=1$ and $\eta_{B}=-1$, and the asterisk
denotes the counter-rotating term. 

From $\rho^{(n)}(t)$, perturbative signals can be determined. The
full perturbative density matrix is given by
\begin{equation}
\kket{\rho^{(n)}(t)}=\left[\sum_{j}\left(K_{j}+K_{j^{*}}+B_{j}+B_{j^{*}}\right)\right]^{n}\kket{\rho^{(0)}},\label{eq:KBexpanded}
\end{equation}
which gives $(4L)^{n}$ different terms, each of which is represented
as a double-sided Feynman diagram. The number of diagrams that must
be calculated can be dramatically reduced when considering the phase
matching or phase cycling conditions in a particular spectrum, which
are sensitive only to some of these contributions to $\rho^{(n)}(t)$.
Further, many calculations are zero in the RWA. Time ordering also
greatly reduces the number of required diagrams when the pulses do
not overlap. There are well established methods to minimize the number
of diagrams required to predict a spectrum, and Ref.~\cite{Rose2020a}
demonstrates how to automate that process.

\begin{figure}
\includegraphics[width=\columnwidth]{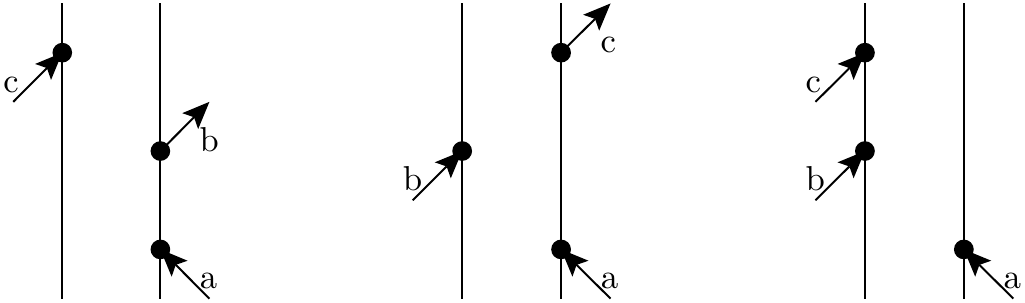}\caption{\label{fig:All-rephasing}Time-ordered Feynman diagrams that contribute
to the rephasing 2DPE, measured in the $\protect\cvec k_{d}=-\protect\cvec k_{a}+\protect\cvec k_{b}+\protect\cvec k_{c}$
direction. Up to 13 other diagrams contribute to the signal when one
or more of the pulses overlap \cite{Rose2020a}.}
\end{figure}

Once the desired diagrams have been determined, the sum in Eq.~\ref{eq:KBexpanded}
can be evaluated with only the relevant diagrams to produce the contributions
to $\ket{\rho^{(n)}}$ that produce the desired signal. For example,
in the case of a phase-matching experiment with detector in the direction
$\cvec k_{d}=\sum_{j}m_{j}\cvec k_{j}$, where $m_{j}$ are integers,
we call the portion of the density matrix that contributes to the
signal $\rho_{k_{d}}^{(n)}$. Then the signal $S_{k_{d}}^{(n)}$ is
calculated using 
\begin{align}
\cvec P_{\mathrm{k}_{d}}^{(n)}(t) & =\langle\boldsymbol{\mu}\rho_{\mathrm{k}_{d}}^{(n)}(t)\rangle\nonumber \\
\tilde{\cvec P}_{\mathrm{k}_{d}}^{(n)}(\omega) & =\int_{-\infty}^{\infty}dte^{i\omega t}\cvec P_{\mathrm{k}_{d}}^{(n)}(t)\nonumber \\
S_{\mathrm{k}_{d}}^{(n)}(\omega) & =\text{Im}\left[\tilde{\varepsilon}_{d}^{*}(\omega)\cvec e_{d}\cdot\tilde{\cvec P}_{\mathrm{k}_{d}}^{(n)}(\omega)\right]\label{eq:S^n}
\end{align}
where $\mathbf{P}_{k_{d}}^{(n)}$ is the $n^{\text{th}}$-order polarization
contributing to the desired signal and the final pulse, with electric
field $\cvec E_{d}$, is the local oscillator used to detect the radiated
field. Figure \ref{fig:All-rephasing} shows the diagrams contributing
to the calculation of the rephasing two-dimensional photon echo (2DPE)
signal when none of the pulses overlap.

The $\alg$ and RKE methods each implement an operation of $O_{j^{(*)}}$
on a density matrix. When they are given a diagram to evaluate, they
compute the required successive $O_{j^{(*)}}$ operations, for example
$B_{c}K_{b}B_{a^{*}}\kket{\rho^{(0)}}$, which is the second diagram
in Fig.~\ref{fig:All-rephasing}.

\subsection{Novel open systems algorithm: $\protect\alg$\label{subsec:UF2}}

We now describe the open systems algorithm we call $\alg$ for the
operators $\{O_{j^{(*)}}\}$, which is an extension of the closed
systems algorithm of the same name presented in Ref.~\onlinecite{Rose2019}.
$\alg$ requires that $\mathscr{L}_{0}$ be time-independent and therefore
that the bath be Markovian. All degrees of freedom corresponding to
non-Markovian effects must be brought into the system, where they
are treated non-perturbatively. With modest computational resource,
we can include several explicit vibrational modes in the system, effectively
giving highly accurate non-Markovian effects to a system that is formally
treated as having a Markovian bath.

We diagonalize $\mathscr{L}_{0}$ by finding the right and left eigenvectors.
The right eigenvectors $\kket{\alpha}$ form a basis and have eigenvalues
$z_{\alpha}$, as
\[
\mathscr{L}_{0}\kket{\alpha}=z_{\alpha}\kket{\alpha}.
\]
The left eigenvectors are defined using overbars as 
\[
\bbra{\bar{\alpha}}\mathscr{L}_{0}=\bbra{\bar{\alpha}}z_{\alpha}.
\]
Since $\mathscr{L}_{0}$ need not be Hermitian, $\kket{\alpha}^{\dagger}\neq\bbra{\bar{\alpha}}$.
We normalize the left and right eigenvectors to satisfy
\begin{equation}
\bbrakket{\bar{\alpha}}{\beta}=\delta_{\alpha\beta}.\label{eq:left-right-norm}
\end{equation}
In the absence of $\mathscr{L}'(t)$, Eq.~\ref{eq:Tdef} gives $\kket{\rho(t)}=\mathcal{T}_{0}(t)\kket{\rho(0)}$.
$\mathcal{T}_{0}$ is diagonal in the basis $\left\{ \kket{\alpha}\right\} $,
so we have
\begin{align}
\kket{\rho(t)} & =\sum_{\alpha}^{N^{2}}e^{z_{\alpha}t}c_{\alpha}\kket{\alpha},\label{eq:rho_of_t}
\end{align}
where $N$ is the dimension of $\mathbb{H}$, which we take to be
finite. If the physical system has an infinite dimensional $\mathbb{H}$,
as in the case of a harmonic oscillator, we truncate $\mathbb{H}$
to dimension $N$, and therefore truncate $\mathscr{L}_{0}$ to dimension
$N^{2}$.

The electric dipole operator acting from the left, $\boldsymbol{\mu}^{K}$,
and from the right, $\boldsymbol{\mu}^{B}$, must be known in the
eigenbasis of $\mathscr{L}_{0}$, where we define matrix elements
\begin{align*}
\boldsymbol{\mu}_{\alpha\beta}^{K} & =\bbra{\bar{\alpha}}\boldsymbol{\mu}^{K}\kket{\beta}.\\
\boldsymbol{\mu}_{\alpha\beta}^{B} & =\bbra{\bar{\alpha}}\boldsymbol{\mu}^{B}\kket{\beta}.
\end{align*}

The derivation of $\alg$ for open systems is formally similar to
that for closed systems in Ref.~\onlinecite{Rose2019}, with replacements
of $U$ by $\mathcal{T}$, the wavefunction $\ket{\psi}$ by the density
vector $\kket{\rho}$, and the dipole operator $\boldsymbol{\mu}$
by $\boldsymbol{\mu}^{K}$ and $\boldsymbol{\mu}^{B}$. Because the
action of the dipole operator on the ket and bra must be considered
separately, the operator $K_{j^{(*)}}$ is joined in the open systems
case by its counterpart $B_{j^{(*)}}$.

We represent $\kket{\rho^{(n)}(t)}$ with coefficients $c_{\alpha}^{(n)}(t)$
that contain only the time dependence induced by the perturbation,
while keeping the evolution due to $\mathscr{L}_{0}$ separate as
\begin{equation}
\kket{\rho^{(n)}(t)}=\sum_{\alpha}e^{z_{\alpha}t}c_{\alpha}^{(n)}(t)\kket{\alpha}.\label{eq:rho-L-eigenbasis}
\end{equation}
With this notation, Eq.~\ref{eq:Odef} gives \begin{widetext}
\begin{align}
O_{j^{(*)}}\kket{\rho^{(n)}(t)} & =\eta_{O}\frac{i}{\hbar}\mathcal{T}_{0}(t)\int_{0}^{\infty}dt'\mathcal{T}_{0}^{-1}(t-t')\sum_{\beta}\kket{\beta}\bbra{\bar{\beta}}\left(\boldsymbol{\mu}^{O}\cdot\cvec e_{j}^{(*)}\varepsilon_{j}^{(*)}(t-t')\right)\sum_{\alpha}e^{z_{\alpha}(t-t')}c_{\alpha}^{(n)}(t-t')\kket{\alpha}\nonumber \\
 & =\eta_{O}\frac{i}{\hbar}\sum_{\beta}e^{z_{\beta}t}\kket{\beta}\int_{-\infty}^{\infty}dt'\theta(t')\underbrace{e^{-z_{\beta}(t-t')}\sum_{\alpha}\left(\boldsymbol{\mu}_{\beta\alpha}^{O}\cdot\cvec e_{j}^{(*)}\varepsilon_{j}^{(*)}(t-t')\right)e^{z_{\alpha}(t-t')}c_{\alpha}^{(n)}(t-t')}_{y_{\beta}(t-t')}.\label{eq:Obar-general}
\end{align}
\end{widetext} The integral in Eq.~\ref{eq:Obar-general} is a convolution,
and we express it in the compact form
\begin{equation}
O_{j^{(*)}}\kket{\rho^{(n)}(t)}=\eta_{O}\frac{i}{\hbar}\sum_{\beta}e^{z_{\beta}t}\kket{\beta}\left[\theta*y_{\beta}\right](t),\label{eq:Obar-compact}
\end{equation}
where 
\[
\left[x*y\right](t)=\int_{-\infty}^{\infty}dt'x(t')y(t-t').
\]
Assuming that $\varepsilon_{j}(t)$ is zero outside the interval $[t_{j,\mathrm{min}},t_{j,\mathrm{max}}]$,
\[
\left[\theta*y_{\beta}\right](t)=\begin{cases}
0 & t<t_{j,\text{min}}\\
r_{\beta}(t) & t_{j,\text{min}}<t<t_{j,\text{max}}\\
C_{\beta} & t>t_{j,\text{max}}
\end{cases}.
\]
for constant $C_{\beta}$. Therefore, we need only calculate this
convolution for $t_{j,\text{min}}<t<t_{j,\text{max}}$.

Physically, we only need to solve for the time dependence due to the
interaction with the pulse while the pulse is nonzero. The rest of
the time dependence is contained in $\mathscr{L}_{0}$ and is therefore
known exactly. This realization drastically reduces the computational
cost of $\alg$ compared to techniques that must use time stepping
for both the system dynamics and the perturbation.

We evaluate the convolution $\left[\theta*y_{\beta}\right](t)$ numerically
to solve for the function $r_{\beta}(t)$ using the FFT and the convolution
theorem. Each electric field envelope $A_{j}(t-t_{j})$ is represented
using $M_{j}$ equally spaced time points, where $M_{j}=(t_{\text{j,max}}-t_{\text{j,min}})/dt_{j}$
and $dt_{j}$ is the spacing between points. Before convolving $y_{\beta}$
is zero-padded up to $2M_{j}-1$ points, and after the convolution
is performed we retrieve only the $M_{j}$ points corresponding to
a linear convolution. Appendix~\ref{subsec:UF2-cost} describes the
computational cost of $\alg$ and shows how it scales with $N$ and
$M$.

\subsection{RKE\label{subsec:RKE}}

The RKE method is an alternative algorithm for evaluating the operators
$\{O_{j^{(*)}}\}$ and is also included in UFSS. It was introduced
in Ref.~\onlinecite{Rose2019} for closed systems. RKE uses the Runga-Kutta
45 (RK45) adaptive time step algorithm to propagate the evolution
due to $\mathscr{L}_{0}$ and a fixed-step Euler method to include
the perturbation $\mathscr{L}'(t)$. It is a direct propagation method,
meaning that it propagates $\kket{\rho}$ forward one step at a time
using the differential form of the equations of motion, Eq.~\ref{eq:EOM}.
RKE is a simple example of a direct-propagation method, and we intend
it to be representative of the computational scaling differences between
$\alg$ and direct-propagation methods; more efficient and higher-order
methods than RKE are possible \cite{Engel1991,Beck2000,Domcke2007,Tsivlin2006,Renziehausen2009,Johansson2012,Fetherolf2017,Yan2017}.

In the absence of pulses, the RK45 method advances the density matrix
$\kket{\rho}$ forward in time according to 
\begin{equation}
\kket{\dot{\rho}}=\mathscr{L}_{0}\kket{\rho}\label{eq:RK45ode}
\end{equation}
where we represent the time evolution due to $\mathscr{L}_{0}$ as
an $N^{2}\times N^{2}$ matrix acting on $\mathbb{L}$, rather than
using $N\times N$ operators on $\mathbb{H}$ as in Eq.~\ref{eq:rhodot}.
We represent a step using the RK45 algorithm alone as $\kket{\rho(t_{i}+dt)}=\mathcal{T}_{0}(dt)\kket{\rho(t_{i})}$.

Starting from $\kket{\rho^{(0)}}$, RKE evaluates diagrams by successive
$O_{j^{(*)}}$ operations. RKE calculates $\kket{\rho_{\beta}}\equiv O_{j^{(*)}}\kket{\rho_{\alpha}}$
for some state $\kket{\rho_{\alpha}}$ as
\begin{align}
\kket{\rho_{\beta}(t_{j,\text{min}}+mdt_{E})} & =\mathcal{T}_{0}(dt_{E})\kket{\rho_{\beta}(t_{j,\text{min}}+(m-1)dt_{E})}\nonumber \\
\mathcal{\phantom{vvv}}+\mathscr{L} & '_{Oj^{(*)}}(t_{j,\text{min}}+mdt_{E})\kket{\rho_{\alpha}(t_{j,\text{min}}+mdt_{E})},\label{eq:Euler}
\end{align}
where we propagate using fixed step size $dt_{E}$ from $t_{j,\text{min}}$
to $t_{j,\text{max}}$. This method accumulates error proportional
to $dt_{E}$. It is possible to construct analogous methods that accumulate
error proportional to $dt_{E}^{2}$ \cite{Renziehausen2009}. Defining
$M_{E}=(t_{j,\text{max}}-t_{j,\text{min}})/dt$, $m$ runs from $0$
to $M_{E}$. Once we obtain $\kket{\rho_{\beta}(t_{j,\text{max}})}$,
the remainder of the time evolution for $t>t_{j,\text{max}}$ is obtained
using the standard RK45 method alone, with a variable time step.

\section{Hamiltonian/Liouvillian generator\label{sec:Vibronic-Model}}

Here we outline the Hamiltonian and Liouvillian generator (HLG) included
as part of UFSS. Note that both $\alg$ and RKE are compatible with
any time-independent Hamiltonian or Liouvillian that can be expressed
as a finite matrix. One need not use the HLG in order to take advantage
of the other modules in UFSS.

HLG is a vibronic model generator, designed to create a Hamiltonian
for a network of $s$ two-level systems (2LS) coupled linearly to
$k$ harmonic vibrational modes. The HLG constructs a Liouvillian
by including coupling of each degree of freedom to a Markovian bath
using either Redfield (full or secular) or diabatic Lindblad formalisms.
Models of this type have been used to describe many systems including
conical intersections in pyrazine and energy transfer in photosynthetic
complexes \cite{Raab2000,Egorova2001,Kleinekathoefer2001,Katz2004,Ishizaki2009,May2011,Caycedo-Soler2012,Killoran2015,Maly2016}.

\subsection{Hamiltonian Structure\label{subsec:Hamiltonian-Structure}}

We begin with an electronic system described by $s$ 2LS,
\[
H_{e}=E_{0}+\sum_{n=1}^{s}E_{n}a_{n}^{\dagger}a_{n}+\sum_{m\neq n}J_{mn}a_{m}^{\dagger}a_{n},
\]
where $a_{n}$ is the annihilation operator for the excited state
in the $n^{th}$ 2LS, $E_{m}$ is the site energy, $E_{0}$ is the
ground state energy, and $J_{mn}$ is a Hermitian matrix of electronic
couplings. The system includes $k$ explicit harmonic vibrational
modes of frequency $\omega_{\alpha}$, generalized momentum $p_{\alpha}$
and coordinate $q_{\alpha}$ with Hamiltonian
\[
H_{ph}=\frac{1}{2}\left(\sum_{\alpha=1}^{k}p_{\alpha}^{2}+\omega_{\alpha}^{2}q_{\alpha}^{2}\right).
\]
We treat standard linear coupling of these modes to the electronic
system as 
\[
H_{e-ph}=\sum_{\alpha=1}^{k}\sum_{n=1}^{s}\omega_{\alpha}^{2}d_{\alpha,n}q_{\alpha}a_{n}^{\dagger}a_{n},
\]
where $d_{\alpha,n}$ indicates the coupling of each vibrational mode
to each 2LS. It is related to the Huang-Rhys factor by 
\[
S_{\alpha,n}=\frac{1}{2}\omega_{\alpha}d_{\alpha,n}^{2}.
\]
The total system Hamiltonian is 
\begin{equation}
H_{0}=H_{e}+H_{ph}+H_{e-ph}.\label{eq:H0}
\end{equation}

If we work in the number basis of the vibrational modes, using the
ladder operators $b_{\alpha}$, with $q_{\alpha}=\frac{1}{\sqrt{2}}(b_{\alpha}+b_{\alpha}^{\dagger}),p_{\alpha}=\frac{i}{\sqrt{2}}(b_{\alpha}-b_{\alpha}^{\dagger})$,
then $H_{0}$ is highly sparse. $H_{e-ph}$ has $2k+1$ entries per
row. $H_{ph}$ is formally infinite in size, so we truncate $H_{0}$
to size $N$ by fixing the total vibrational occupation number. Note
that Eq.~\ref{eq:H0} is block diagonal with $s+1$ blocks. Each
of these blocks is an optically separated manifold, and we index manifolds
using $X$ and $Y$, where $X,Y$ can refer to the ground-state manifold
(GSM), the singly excited manifold (SEM), the doubly excited manifold
(DEM), and so on. Each block has a size $N_{X}$, and $N=N_{GSM}+N_{SEM}+N_{DEM}+...$.
The block diagonal form of $H_{0}$ is not required by $\alg$, but
it allows useful simplifications in certain cases, which are discussed
briefly at the end of this section and in Appendix~\ref{sec:Computational-cost}
.

\subsection{Liouvillian Structure\label{subsec:Liouvillian-Structure}}

Using the Hamiltonian from Eq.~\ref{eq:H0}, we construct the unitary
part of the Liouvillian, $\mathscr{L}_{U}=\mathscr{L}_{0}-i\hbar D$,
where $\mathscr{L}_{0}$ is defined in Eq.~\ref{eq:L0}. UFSS allows
coupling of a Markovian bath to all degrees of freedom of $H_{0}$
using either the Redfield \cite{Breuer2002} or Lindblad formalisms
\cite{Gardiner2004} or a user-specified combination of them, should
that be desirable.

\subsubsection{Redfield}

Redfield theory arises from microscopic derivations of the properties
of the bath and system-bath coupling, in contrast to diabatic Lindblad
theory, which requires phenomenological relaxation and dephasing parameters.
Given a system-bath coupling Hamiltonian of the form 
\[
H_{SB}=\sum_{r}O_{r,S}\otimes O_{r,B},
\]
where $O_{r,S}$ is an operator defined in the Hilbert space of the
system, and $O_{r,B}$ is an operator defined in the Hilbert space
of the bath, the dissipation tensor is defined as 
\[
Y_{ijkl}(O_{r,S})=\sum_{r,r'}\bra iO_{r,S}\ket k\bra lO_{r',S}\ket jC_{rr'}(\omega_{ki}),
\]
where $H_{0}\ket i=\hbar\omega_{i}\ket i$ defines the eigenvectors
of $H_{0}$, given by Eq.~\ref{eq:H0}, and
\[
C_{rr'}(t)=\langle O_{r,B}(t)O_{r',B}(0)\rangle
\]
are the two-point correlation functions of the phonon modes of the
bath. The index $r$ specifies either the site index $n$ or the vibrational
mode index $\alpha$. We assume that the cross-correlation terms are
zero, and so $C_{rr'}(t)=\delta_{rr'}C_{rr'}(t)$. The Fourier transform
of $C(t)$ is specified using a spectral density $J(\omega)$ as
\begin{align*}
\Re\left[C(\omega)\right] & =\frac{1}{2}\hbar J(\omega)\coth\left(\frac{\hbar\omega}{2k_{B}T}\right)\\
\Im\left[C(\omega)\right] & =\frac{1}{\pi}P\int_{-\infty}^{\infty}d\omega'\frac{\Re\left[C(\omega')\right]}{\omega-\omega'},
\end{align*}
where $P$ indicates the Cauchy principal value. In the examples that
follow, we use an Ohmic spectral density with the Drude-Lorentz cut-off
function
\[
J(\omega)=2\omega\lambda\frac{\gamma}{\omega^{2}+\gamma^{2}}
\]
where $\lambda$ is the strength of the system-bath coupling and $\gamma$
is the cut-off frequency of the bath.Users are also free to specify
any spectral density function that is appropriate to their system.

The Redfield tensor is 
\[
\begin{split}
R_{ijkl}(O)=&-\left(Y_{ijkl}(O)+Y_{jilk}^{*}(O)\right)+\delta_{jl}\sum_{\sigma}Y_{\sigma ik\sigma}(O)\\
&+\delta_{ik}\sum_{\sigma}Y_{\sigma jl\sigma}^{*}(O)
\end{split}
\]

We consider system-bath coupling to each site $n$ via the Redfield
tensor $R(a_{n}^{\dagger}a_{n})$, and coupling to each vibrational
mode via the Redfield tensor $R(q_{\alpha})$, where $q_{\alpha}=\frac{1}{\sqrt{2}}\left(b_{\alpha}^{\dagger}+b_{\alpha}\right)$,
so that the dissipative part of $\mathscr{L}_{0}$ (see Eq.~\ref{eq:L0})
is 
\[
D=-R(a_{n}^{\dagger}a_{n})-R(q_{\alpha}).
\]

We also optionally include in $D$ the relaxation of electronic excitations
to the ground state via the Redfield tensor $R(a_{n}^{\dagger}+a_{n})$.
The microscopic derivation of the two-time correlation function $C(t)$
that mediates such relaxation processes is more difficult, as it involves
non-adiabatic coupling terms \cite{Brueggemann2003} and is rarely
used. As such, by default we have included a flat spectral density
for inter-manifold relaxation processes such that the associate $C(\omega)$
is 
\[
C(\omega)=\begin{cases}
\gamma_{r} & \omega>0\\
\gamma_{r}e^{\hbar\omega/k_{B}T} & \omega<0
\end{cases},
\]
where $\gamma_{r}$ provides a phenomenological inter-manifold relaxation
rate. In the secular approximation (see below), this phenomenological
form reduces to a commonly used approach \cite{May2011,Suess2019,Maly2020}.
Users are free to provide more complicated spectral densities to describe
this type of process.

The commonly used secular approximation sets $R_{ijkl}=0,$ except
when $|\omega_{ij}-\omega_{kl}|=0$, which guarantees positivity of
the density matrix. The remaining simple form of $\mathscr{L}_{0}$
only couples the populations of the density matrix couple to one another
(see Appendix~\ref{sec:Computational-cost} for exceptions)\cite{Breuer2002,Ishizaki2009,Maly2016}.
In this case, $\mathscr{L}_{0}$ consists of a single $N\times N$
block for the populations and is otherwise diagonal. This simplification
has important implications on the computational complexity of both
the $\alg$ and RKE algorithms. In particular, the cost of diagonalizing
$\mathscr{L}_{0}$ becomes $\sim N^{3}$, which is the same scaling
as the cost of diagonalizing $H_{0}$, and in stark contrast the the
cost of diagonalizing a generic $\mathscr{L}_{0}$, which is $\sim N^{6}$.

\subsubsection{Diabatic Lindblad}

The Lindblad formalism is widely used to describe open quantum systems
using a small number of phenomenological dephasing and relaxation
constants. It guarantees the complete positivity of $\mathscr{L}_{0}$.
Secular Redfield theory can be mapped to a Lindblad structure using
operators in the eigenbasis of the Hamiltonian. Recently it has been
shown that full Redfield theory can be mapped to a Lindblad structure
in some cases (again this mapping is done in the eigenbasis of the
Hamiltonian) \cite{McCauley2020}. Here, we present a Lindblad formalism
in the diabatic basis, which can be fairly accurate for short time-scales,
though it does not produce a thermal distribution at infinite time,
so must break down for long time-scales \cite{Egorova2001,Kleinekathoefer2001}.

If $O$ is an operator on $\mathbb{H}$, the Lindblad superoperator
is
\[
L[O]\rho=O\rho O^{\dagger}-\frac{1}{2}\left(O^{\dagger}O\rho+\rho O^{\dagger}O\right).
\]
We consider dissipation in vibrational modes and both inter- and intra-manifold
dephasing and relaxation of the electronic modes.

For vibrational mode $\alpha$, we describe coupling to the rest of
the bath with the dissipation operator
\begin{equation}
D_{v}=\sum_{\alpha}\gamma_{v,\alpha}\left(N_{\text{th}}L[b_{\alpha}^{\dagger}]+(N_{\text{th}}+1)L[b_{\alpha}]\right),\label{eq:Dmonomer}
\end{equation}
where $\gamma_{v,\alpha}$ is the thermalization rate of the $\alpha^{\text{th}}$
mode and $N_{\text{th}}=\langle b_{\alpha}^{\dagger}b_{\alpha}\rangle=1/(\exp(\beta\hbar\omega_{\alpha})-1)$
is the average number of excitations at equilibrium of a mode with
energy $\hbar\omega_{\alpha}$ coupled to a Markovian bath of temperature
$T$, with $\beta=1/k_{B}T$, with $k_{B}$ the Boltzmann constant
\cite{Gardiner2004}.

We describe inter-manifold electronic relaxation and the complementary
incoherent thermal excitation processes with 
\[
D_{r1}=\sum_{n=1}^{s}\gamma_{r1,n}\left(C_{gn}L[a_{n}]+C_{ng}L[a_{n}^{\dagger}]\right),
\]
where $C_{nm}=\frac{e^{-\beta E_{n}}}{e^{-\beta E_{n}}+e^{-\beta E_{m}}}$.
Intra-manifold relaxation processes are described by
\[
D_{r2}=\sum_{m\neq n}\gamma_{r2,nm}C_{nm}L[a_{n}^{\dagger}a_{m}].
\]
Since $C_{nm}/C_{mn}=e^{-\beta(E_{n}-E_{m})}$, we obtain a thermal
distribution of eigenstates as $t\rightarrow\infty$ if $J_{mn}=0$.

Relaxation processes necessarily give rise to dephasing. We include
additional intra-manifold dephasing using 
\[
D_{d2}=\sum_{n\neq m}\gamma_{d2,nm}L\left[a_{n}^{\dagger}a_{n}-a_{m}^{\dagger}a_{m}\right].
\]
All of the above processes also give rise to inter-manifold (optical)
dephasing. We include additional, pure inter-manifold dephasing using
\begin{equation}
D_{d1}=\gamma_{d1}L\left[\sum_{n=1}^{s}a_{n}^{\dagger}a_{n}\right].\label{eq:inter-manifold-deph}
\end{equation}
In the common case where $\gamma_{d1}$ is larger than all the other
bath coupling rates, the homogeneous linewidth(s) are dominated by
the $D_{d1}$ term. Putting all of these operators together we arrive
at the total dissipation operator 
\begin{equation}
D=D_{r1}+D_{r2}+D_{d1}+D_{d2}+D_{\nu}.\label{eq:D-total}
\end{equation}

When modeling optical spectroscopies, $H_{0}$ is often block diagonal,
and therefore the system is composed of distinct manifolds, as constructed
in Sec.~\ref{subsec:Hamiltonian-Structure}. In the case where we
can also neglect inter-manifold relaxation processes ($\gamma_{r1,n}=0$),
the total Liouvillian $\mathscr{L}_{0}$ is also block diagonal. Under
these assumptions $\mathscr{L}_{0}$ can be arranged into blocks of
size $N_{X}^{2}\times N_{Y}^{2}$, allowing us to save both computational
cost and memory, both for diagonalization (if applicable) as well
as for use with $\alg$ or RKE. The cost of diagonalizing $\mathscr{L}_{0}$
is then $\sim N_{X}^{3}N_{Y}^{3}$ .

\section{Computational advantage\label{sec:Computational-advantage}}

We compare the computational costs of the two UFSS propagation methods,
$\alg$ and RKE, for three methods of including the bath: secular
Redfield, full Redfield, and diabatic Lindblad. We show that the convolution-based
$\alg$ is over 200 times faster than the direct-propagation RKE method
for small systems. Asymptotically the relative performance of $\alg$
depends strongly on the method of treating the bath. Appendix~\ref{sec:Computational-cost}
derives the asymptotic computational complexity of these methods,
and the results are summarized in Table~\ref{tab:Computational-cost-summary}.
These results predict that the $\alg$ method is always more efficient
than the direct propagation method for secular Redfield. While these
computational complexity results do not include the memory requirements
of the algorithms, we demonstrate below that for full Redfield, $\alg$
is more efficient than RKE up until system sizes where $\mathscr{L}_{0}$
requires at least 20 GB to store, as summarized in the last column
of the table.

\begin{table}
\caption{\label{tab:Computational-cost-summary} Scaling of computational cost
with the Hamiltonian dimension $N$. \textquotedblright$\mathscr{L}_{0}$
Diag\textquotedblleft{} is the diagonalization of $\mathscr{L}_{0}$.
Derivations are in Appendix~\ref{sec:Computational-cost}.}
\begin{tabular}{ccccc}
 &  &  &  & \tabularnewline
\hline 
\hline 
 & $\alg$ & \multicolumn{1}{c}{RKE} & $\mathscr{L}_{0}$ Diag. & $\alg$ advantage\tabularnewline
\hline 
Full Redfield & $N^{4}$ & $N^{4}$ & $N^{6}$ & up to large $N$\tabularnewline
Secular Redfield & $N^{3}$ & $N^{3}$ & $N^{3}$ & all $N$\tabularnewline
Diabatic Lindblad & $N^{4}$ & $N^{2}$ & $N^{6}$ & $N\lesssim100$\tabularnewline
\hline 
 &  &  &  & \tabularnewline
\end{tabular}
\end{table}

$\alg$ has a one-time cost of diagonalizing $\mathscr{L}_{0}$. While
$\mathscr{L}_{0}$ is generally a matrix of size $N^{2}\times N^{2}$
with diagonalization cost $\sim N^{6}$, in the secular approximation
this cost is only $\sim N^{3}$, as explained in Sec.~\ref{sec:Computational-cost}.
Despite diagonalization being an expensive calculation for full Redfield
and the diabatic Lindblad models, it does not necessarily contribute
significantly to the total cost of calculating spectra, because the
diagonal form is reused for each set of pulse delays, pulse shapes,
pulse polarizations, etc. Consider a sample 2DPE signal $S^{(3)}(\tau,T,\omega_{t})$
with 100 coherence times $\tau$ and 20 population times $T$, as
well as a sample TA signal $S^{(3)}(T,\omega_{t})$ with 100 delay
times. Since the 2DPE spectrum requires 3-16 diagrams evaluated with
2000 different pulse delays, the cost of the diagonalization is effectively
amortized over $>6000$ calculations. The TA calculation requires
6-16 diagrams evaluated at only 100 delay times, so amortizes the
diagonalization cost over $\approx600$ calculations. The RKE method
does not require diagonalization, so the system size at which it becomes
cost effective to use the RKE method in principle depends on what
type of spectrum is being considered. In diabatic Lindblad, diagonalization
cost begins to be limiting for TA spectra around the same size that
RKE becomes more efficient, regardless.

\begin{figure}
\includegraphics[width=\columnwidth]{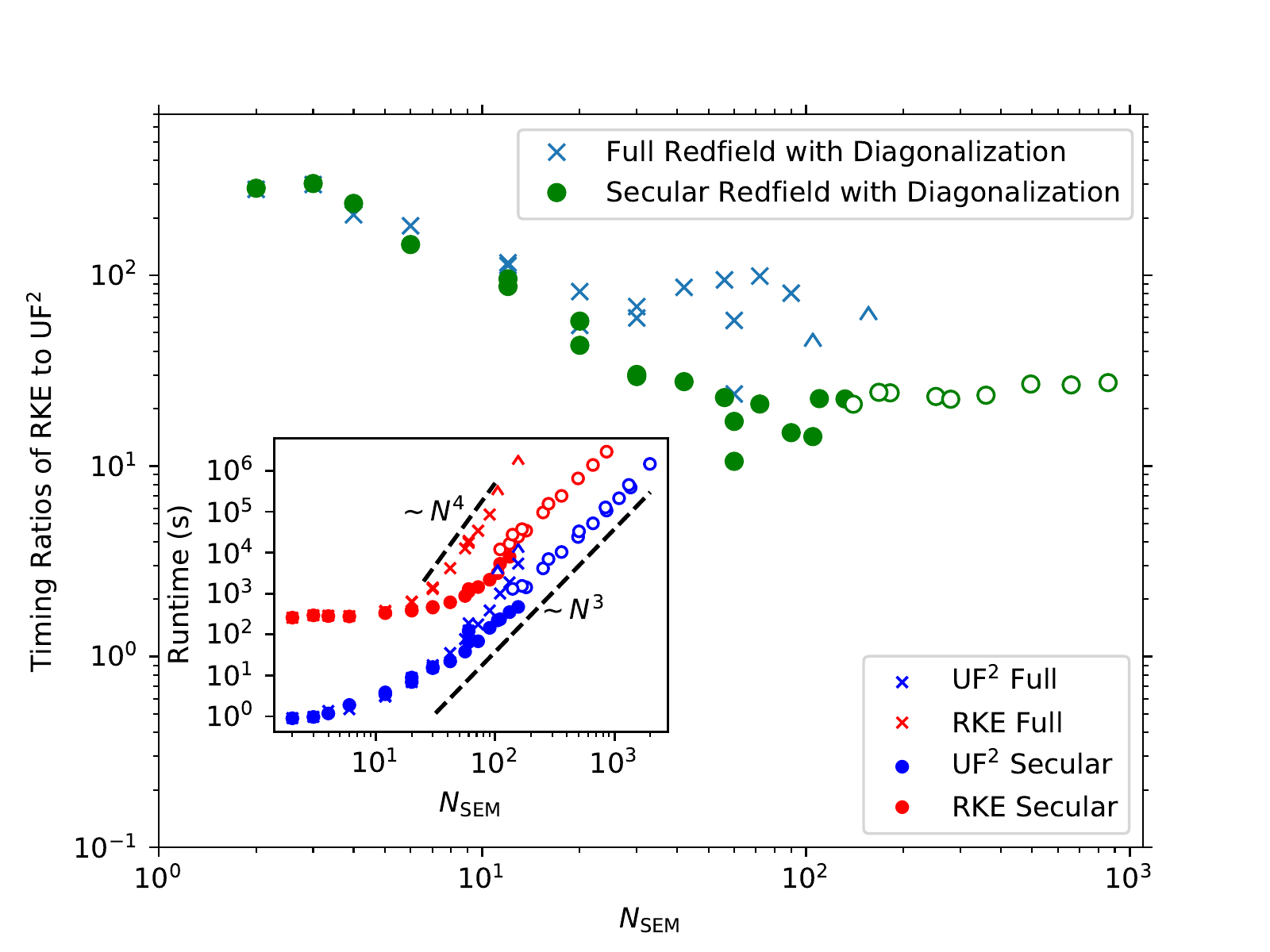}\caption{\label{fig:Redfield-timing-ratios}Timing ratios of RKE to $\protect\alg$
for TA spectra with 100 time delays using the Redfield formalism for
a range of systems and parameters (closed circles and x's). For large
systems, we simulate a single time delay (open circles and carats)
with only one ESA diagram and multiply the runtime by 600 to effectively
treat 100 delay times and 6 diagrams, due to the long runtimes. $N_{SEM}$
is the dimension of the Hamiltonian describing the truncated singly
excited manifold, large enough to converge the spectra. Ratios include
the cost of diagonalizing $\mathscr{L}_{0}$ in the $\protect\alg$
costs. The cost of diagonalization is never important in the secular
approximation, and the $\protect\alg$ advantage plateaus at a factor
of 20 for large $N_{SEM}$, matching the predictions from Appendix~\ref{sec:Computational-cost}.
With full Redfield, ignoring diagonalization, asymptotically $\protect\alg$
has a theoretical relative advantage of 40, in good agreement with
the results (including diagonalization costs) shown here. Inset shows
the associated runtimes without diagonalization costs. Dashed lines
show the predicted asymptotic scalings from Appendix~\ref{sec:Computational-cost}.}
\end{figure}

\begin{figure}
\includegraphics[width=\columnwidth]{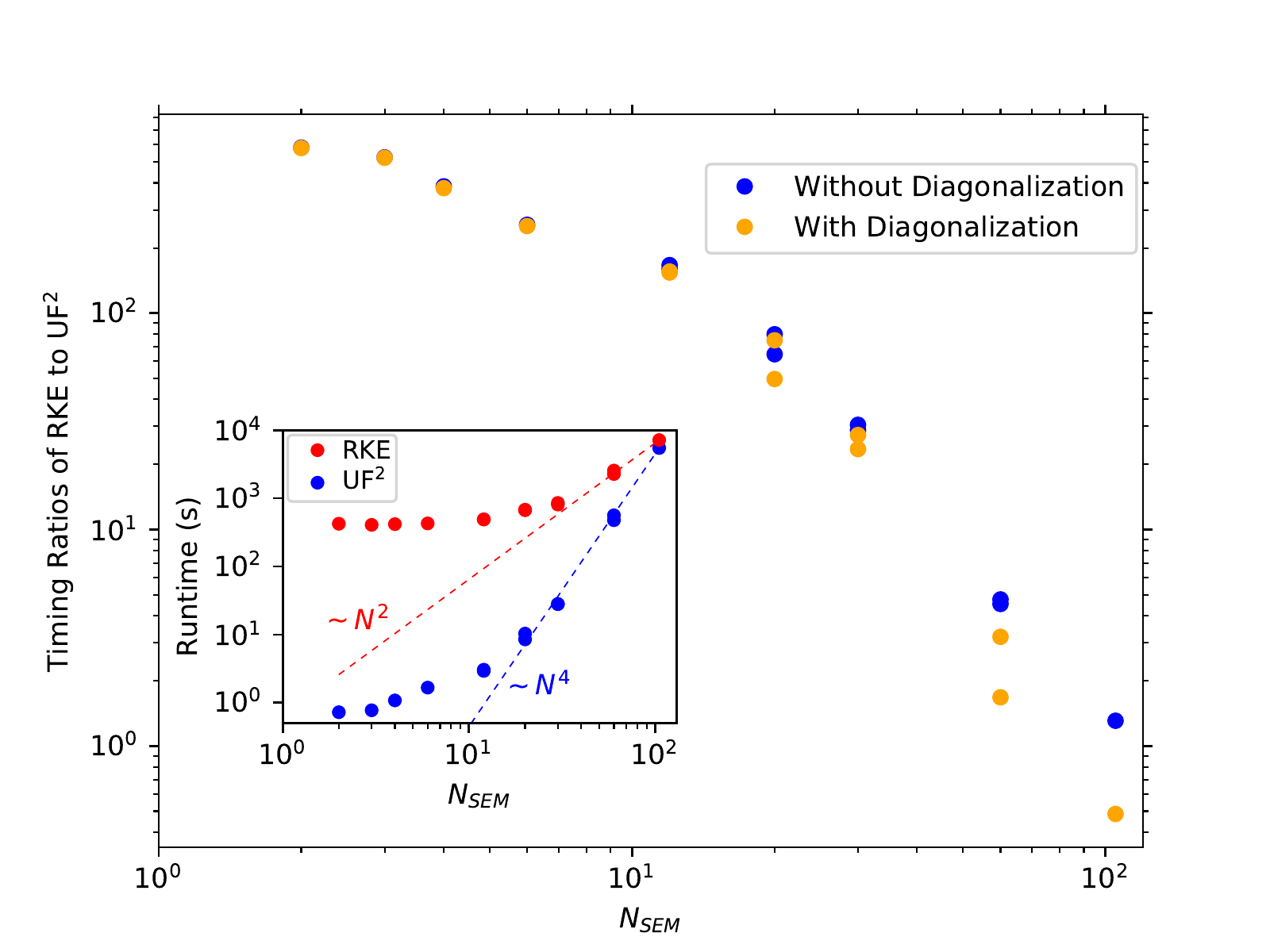}\caption{\label{fig:Timing-ratios}Timing ratios of RKE to $\protect\alg$
methods for TA spectra with 100 time delays for a range of systems
and parameters, described in the text. $N_{SEM}$ is the dimension
of the Hamiltonian describing the truncated singly excited manifold,
large enough to converge the spectra. Blue points show the timing
ratios for the evaluation of the spectra but not the diagonalization
of $\mathscr{L}_{0}$. Orange points include the cost of the diagonalization,
which is insignificant with small $N_{SEM}$. Near $N_{SEM}=100$,
the direct-propagation RKE method becomes more efficient than $\protect\alg$.
Inset shows the time required to calculate the TA spectrum using RKE
and $\protect\alg$, without the diagonalization cost included. The
dashed lines show the slopes of the predicted asymptotic scaling of
each algorithm.}
\end{figure}

The computational costs of predicting spectra depend upon both the
size and structure of $\mathscr{L}_{0}$. To make concrete comparisons
between $\alg$ and RKE, we use a vibronic Hamiltonian $H_{0}$ coupled
to a Markovian bath, as outlined in Section~\ref{sec:Vibronic-Model}.
Figures~\ref{fig:Redfield-timing-ratios} and \ref{fig:Timing-ratios}
show the ratio of the computation time of RKE to the computation time
of $\alg$ for TA spectra with Redfield and Lindblad models, respectively.
We consider systems with the number of sites and number of vibrational
modes equal ($s=k$) and varying from 2 to 4. The energy scale of
the problem is defined by the $k$ nearly identical vibrational frequencies
$\omega_{\alpha}$, which are all within about 1\% of $\omega_{0}$.
The modes are detuned for convenience, to avoid degeneracies in the
ground state manifold, which makes the structure for the Redfield
tensor simpler in the secular approximation. We use the RWA and, after
rotating away the optical gap, the site energies $E_{i}$ vary from
$0-1.5\omega_{0}^{-1}$ , and the coupling terms $J_{mn}$ vary from
$0-0.5\omega_{0}^{-1}$. We use the same value $d_{\alpha,n}=d\delta_{\alpha,n}$
for each $\alpha,n$ pair, where $\delta_{i,j}$ is the Kronecker
delta, and choose values of $d$ from $0$ to $1.5$. Larger values
of $d$ require a larger truncation size $N$ for the spectra to converge.
For Redfield theory we use the same bath parameters for both the sites
and the vibrational modes, $\lambda=0.05\omega_{0}$ and $\gamma=\omega_{0}$,
and have taken $\Im[C(\omega)]=0$. In the diabatic Lindblad model,
we include a Markovian bath using $\gamma_{r2,i}=0.05\omega_{0}$,
$\gamma_{\nu,\alpha}=0.05\omega_{0}$ and $\gamma_{d1}=0.2\omega_{0}$
(with $\gamma_{r1}=\gamma_{d2}=0$). The optical pulses have Gaussian
envelopes with standard deviation $\sigma=\omega_{0}^{-1}$, centered
on the transition $E_{1}-E_{0}$. All sites $i$ have parallel dipole
moments, with magnitudes varying from 0.7 to 1, and we use the Condon
approximation that $\mu$ is independent of vibrational coordinate.
We choose the number of vibrational states in the simulations to be
sufficiently large by using $\alg$ to generate a TA signal and seeking
the truncation size $N$ that converges the resulting spectra within
1\% using an $\ell_{2}$ norm over the full spectrum. We perform this
convergence separately for each case of $s,d$. For each choice of
$s,$$d$, we use the same $N$ for the RKE calculations. The optical
field parameters ($M$ and $dt$ for $\alg$ and $M_{E}$ and $dt_{E}$
for RKE) were determined by testing on some of the smaller systems
and were held constant for all $s,d$. Values of $t$ were selected
in order to resolve all optical oscillation frequencies in the RWA
and to resolve the homogeneous linewidth. We took the inter-manifold
relaxation rate $\gamma_{r1}=0$, so that the optical manifolds are
separable and $\mathscr{L}_{0}$ is block diagonal in all cases. All
calculations were performed on an Intel Xeon E5-2640 v4 CPU with a
2.40 GHz clock speed and 96 GB of RAM.

$\alg$ always outperforms RKE for secular Redfield theory, saturating
at 20 times faster for large $N_{SEM}$, as predicted in Appendix
\ref{sec:Computational-cost}. In all three formalisms, $\alg$ is
200-500 times faster than RKE for small $N_{SEM}$. For full Redfield,
with large $N_{SEM}$, the cost of diagonalization should eventually
cause RKE to outperform $\alg$, but we see that $\alg$ is approximately
100 times faster even at the larger system sizes studied. The memory
requirements of constructing $\mathscr{L}_{0}$ using full Redfield
theory become limiting, as $\mathscr{L}_{0}$ is a full $N^{2}\times N^{2}$
matrix of complex floats. The largest $\mathscr{L}_{0}$ studied was
20 GB. For diabatic Lindblad the memory requirements of diagonalizing
$\mathscr{L}_{0}$ are similar.

The insets of both figures show the wall-clock runtimes for each value
of $N_{SEM}$, not including the diagonalization cost for $\alg$.
The dashed lines in the insets show that as $N_{SEM}$ increases,
the expected asymptotic scalings from Table~\ref{tab:Computational-cost-summary}
are obeyed. For the diabatic Lindblad model, $\alg$ is more efficient
than RKE with sufficiently small systems, the superior cost scaling
of RKE leads to a crossover in runtimes near $N_{SEM}=100$, corresponding
to $\mathscr{L}_{0}$ having blocks of dimension $N_{SEM}^{2}=10^{4}$.
This crossover is consistent with our result with closed systems in
Ref.~\onlinecite{Rose2019}, where the $\alg$ method was more efficient
than RKE for $N_{SEM}$ smaller than $10^{4}-10^{5}$. For small $N_{SEM}$,
the cost of diagonalization is negligible, as shown by the blue and
orange dots in Figure \ref{fig:Timing-ratios} all overlapping for
$N_{SEM}<10$. As $N_{SEM}$ increases, the cost of diagonalization
becomes apparent as the colors separate. With or without diagonalization,
Fig.~\ref{fig:Timing-ratios} shows that the crossover occurs with
$N_{SEM}\approx100$.

\section{Comparison of $\protect\alg$ to analytic results \label{sec:Comparisons-to-literature}}

We now demonstrate that the signals produced by $\alg$ reproduce
an analytical solution for the rephasing 2D photon echo (2DPE) signal
for the optical Bloch equations using Gaussian pulses \cite{Smallwood2017}.
Reference~\onlinecite{Smallwood2017} considered a small ($N=3$)
system that can be mapped to the Hamiltonian described by $s=2$ and
$k=0$ from Sec.~\ref{sec:Vibronic-Model}, with the doubly excited
state removed. Although this is a small system, these comparisons
are some of the only available analytical solutions including finite
pulse durations, and thus provides a benchmark to show that $\alg$
calculates spectra with a high degree of accuracy. $\alg$ converges
to within 1\% of the analytical result using just $M=25$ points to
discretize $\text{\ensuremath{\varepsilon}}_{j}(t)$.

In this model, the energy difference between the two excited states
is $\hbar\omega_{0}=E_{2}-E_{1}$. All pulses are taken to have identical
Gaussian envelopes
\[
A(t)=\frac{1}{\sqrt{2\pi}\sigma}e^{-t^{2}/2\sigma^{2}},
\]
where $\sigma=\omega_{0}^{-1}$ and have central frequency $\omega_{j}=(E_{2}+E_{1})/(2\hbar)$.
In the RWA, we are free to set $\omega_{j}=0$ for all pulses, which
we do. The model includes phenomenological dephasing rates and population
decay rates of $0.2\omega_{0}$ and $0.1\omega_{0}$, respectively.
This bath coupling is similar to a Lindblad formalism like the one
outlined in Sec.~\ref{subsec:Liouvillian-Structure}, except that
it does not conserve the total probability of the density matrix.
Rather than using the HLG included in UFSS, for this comparison we
separately created the model described in Ref.~\onlinecite{Smallwood2017}.
The construction of $\mathscr{L}_{0}$ and the evaluation of the resulting
spectra is demonstrated in the Jupyter notebook $\texttt{Smallwood2017Comparison.ipynb}$,
available in the UFSS repository.

\begin{figure}
\includegraphics[width=\columnwidth]{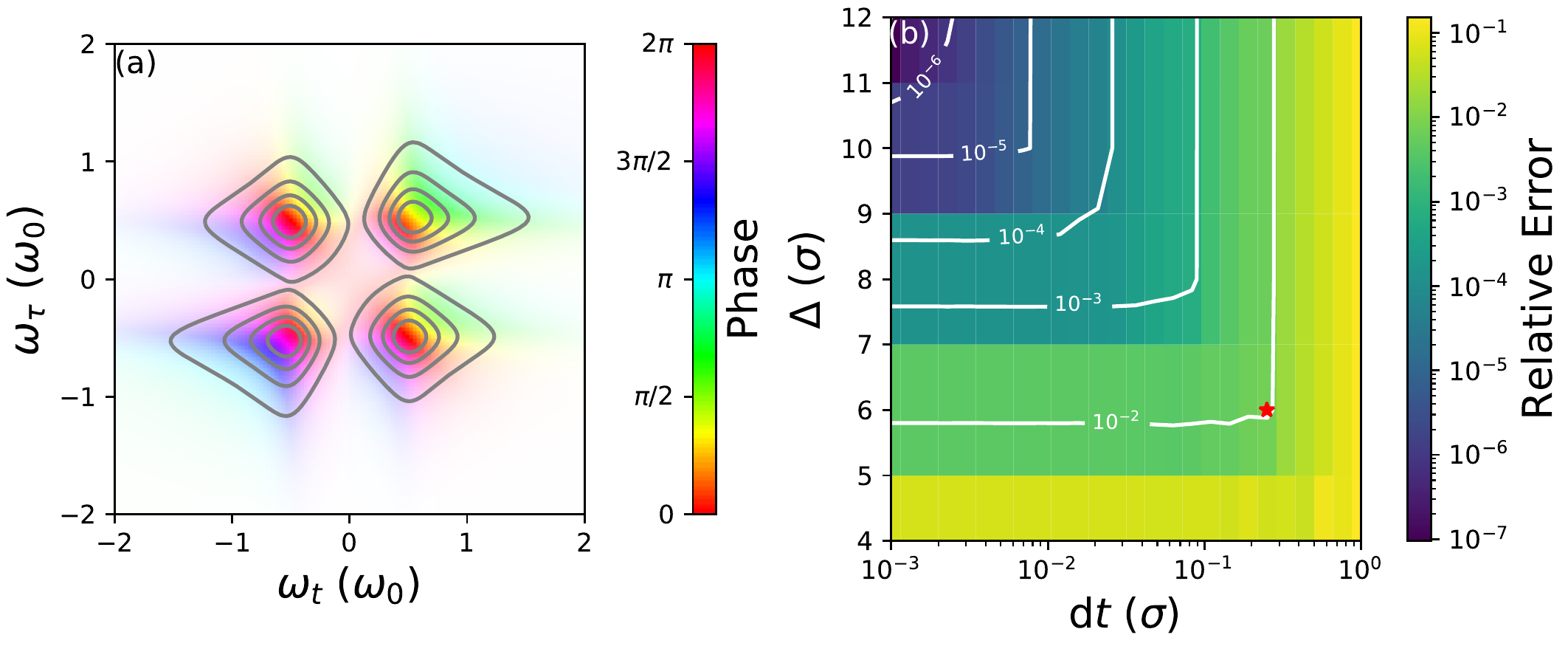}
\caption{\label{fig:Smallwood-comparison}(a) Reproduction of Smallwood's Figure
3e) using their analytical forms. The color shows the phase of the
complex signal $P^{(3)}(\omega_{\tau},T,\omega_{t})$, while the intensity
of the color shows the magnitude of the signal. The numerical result
using $\protect\alg$ appears visually identical so is not shown.
(b) $\ell_{2}$ norm of the difference between the analytical solution
shown in (a) and the result from $\protect\alg$, both sampled on
a mesh of $801\times801$ $\omega_{t}$, $\omega_{\tau}$ points,
as a function of $dt$ and the pulse duration $\Delta$ used in the
numerical convolutions of Eq.~\ref{eq:Obar-compact}. We evaluate
$P^{(3)}(\tau,T,t)$ for $t$ and $\tau$ ranging from $-100\sigma$
to $100\sigma$ in steps of $0.25\sigma$. The red star in (b) indicates
the smallest value of $M=\Delta/dt+1$ needed to reach 1\% agreement
with the analytical solution shown in (a), and corresponds to $M=25$
points. Note that for $\Delta=12\sigma$, the figure shows that $\protect\alg$
converges to the analytical signal as $dt^{2}$.}
\end{figure}

The 2DPE signal is the result of three pulses, which gives rise to
an emitted field $P(\tau,T,t)$. $\alg$ is designed to calculate
$P(\tau,T,t)$, while the result from Ref.~\onlinecite{Smallwood2017}
is for the quantity 
\[
P(\omega_{\tau},T,\omega_{t})=\frac{1}{2\pi}\int_{-\infty}^{\infty}dte^{i\omega_{t}t}\int_{-\infty}^{\infty}d\tau e^{-i\omega_{\tau}\tau}P(\tau,T,t).
\]
To compare $\alg$ to the analytical solutions, we calculate $P(\tau,T,t)$
for a discrete set of $\tau,T,t$, and take a 2D discrete Fourier
transform with respect to $\tau$ and $t$. Since $A(t)$ is symmetric,
we evaluate $A(t-t_{j})$ on the interval $t\in[t_{j,\text{min}},t_{j,\text{max}}]$
with spacing $dt$, and choose $t_{\text{j,max}}-t_{j}=t_{j}-t_{\text{min}}$,
where $t_{j}$ is the arrival time of the $j^{th}$ pulse. For symmetric
pulses, $\alg$ converges most rapidly when the value $t-t_{j}=0$
is included at the center of the discretization interval and the endpoints
of the interval are also included. We define the duration of the pulse,
$\Delta_{j}\equiv t_{j,\text{max}}-t_{j,\text{min}}$ and the number
of points $M_{j}=\Delta_{j}/dt+1$. All the pulses are identical,
and so we use $M=M_{j}$, and therefore $\Delta=\Delta_{j}$ for each
pulse.

Figure \ref{fig:Smallwood-comparison}(a) shows the analytical result
for $P(\omega_{\tau},0,\omega_{t})$, which provides a benchmark for
quantifying the convergence behavior of $\alg$. Fig.~\ref{fig:Smallwood-comparison}(b)
shows the $\ell_{2}$ norm of the difference between the analytical
solution shown in (a) and the result from $\alg$. The white contours
in Fig.~\ref{fig:Smallwood-comparison}(b) show that the errors due
to $\Delta$ and $dt$ are nearly independent, since the contours
are approximately composed of horizontal and vertical lines, leading
to the appearance of terraces in the color plot. Inspection of the
top of that plot shows that $\alg$ converges to the analytical signal
as $dt^{2}$, when $\Delta$ is sufficiently large that only the error
from $dt$ is significant. $\alg$ reproduces the analytical result
to within 1\% by using $\Delta=6\sigma$ and $dt=0.25\sigma$, corresponding
to $M=25$. The spectra attained using these parameters is not shown,
as it is visually identical to Fig.~\ref{fig:Smallwood-comparison}(a).
We conclude that $\alg$ accurately predicts nonlinear optical spectra
including finite-pulse duration effects in systems with small $M$.

\section{Conclusion}

We have presented three separate components of the Ultrafast Spectroscopy
Suite (UFSS), which is a modular suite of tools designed for predicting
nonlinear optical spectra. We have presented a novel algorithm called
$\alg$ that uses the convolution theorem to efficiently propagate
the time evolution of eigenstates of the system Liouvillian. $\alg$
is designed for evaluating the contributions to spectroscopic signals
from the Feynman diagrams that organize perturbative calculations
of nonlinear optical spectra. $\alg$ is the open-systems extension
of the closed-system algorithm of the same name presented in Ref.~\onlinecite{Rose2019}.
We have also presented a direct propagation technique called RKE and
a Hamiltonian/Liouvillian generator (HLG), which creates Hamiltonians
and Liouvillians for vibronic systems coupled to a Markovian bath.
Using the HLG, we have demonstrated that $\alg$ can be over 500 times
faster than RKE for systems with small Hilbert space dimension $N$.
Using a secular Redfield model, $\alg$ is always faster than RKE,
and with full Redfield $\alg$ is faster up to system sizes where
$\mathscr{L}_{0}$ rqeuires more than 20 GB of memory. In the diabatic
Lindblad model, $\alg$ outperforms RKE for $N\lesssim100$. Both
$\alg$ and RKE methods are available with UFSS and can be used where
appropriate.

A fourth module of UFSS, called the diagram generator (DG), is presented
in Ref.~\cite{Rose2020a}. The DG automatically generates all of
the necessary Feynman diagrams that are needed to calculate a spectroscopic
signal given the phase-matching (or phase-cycling) condition, the
pulse shapes and pulse arrival times. Taken all together, the UFSS
allows fast and automated calculations of nonlinear optical spectra
of any perturbative order, for arbitrary pulse shapes, since $\alg$
and RKE can both automatically calculate spectra given a list of Feynman
diagrams. If desired, a user of UFSS need not concern themselves with
the details of the perturbative calculations carried out by $\alg$
and RKE. They simply must input the phase-matching conditions and
pulse shapes of interest.

Each module of UFSS presented here can also be used separately. $\alg$
and RKE can calculate the signal due to only a single Feynman diagram,
or only the time-ordered diagrams, as is done when comparing to the
analytical results of Ref.~\onlinecite{Smallwood2017}. $\alg$ and
RKE are compatible with any Hamiltonian or Liouvillian that is time-independent
and can be expressed as or well-approximated by a finite matrix. Thus
users are free to input their own model systems, and we include helper
functions for saving other Hamiltonians and Liouvillians into a format
compatible with $\alg$ and RKE.

UFSS is available under the MIT license at \href{https://github.com/peterarose/ufss}{github}.
The repository includes Jupyter notebooks that generate Figures~\ref{fig:All-rephasing}
and \ref{fig:Smallwood-comparison} from this manuscript and scripts
that generate Figures~\ref{fig:Redfield-timing-ratios} and \ref{fig:Timing-ratios}.
\begin{acknowledgments}
We thank an anonymous reviewer for the suggestion to consider the
Redfield methods and acknowledge support from the Natural Sciences
and Engineering Research Council of Canada (NSERC) and the Ontario
Trillium Scholarship.
\end{acknowledgments}

\subsection*{Data Availability Statement}

The data that support the findings of this study are available from
the corresponding author upon reasonable request. In addition, the
code to generate all of the figures in this manuscript are available
at \href{https://github.com/peterarose/ufss}{github}.

\appendix

\section{Computational cost\label{sec:Computational-cost}}

Here we derive the asymptotic computational costs of the $\alg$ and
RKE methods for the secular Redfield, full Redfield, and diabatic
Lindblad models. For an arbitrary $n^{\text{th}}$-order spectroscopy,
both $\alg$ and RKE must calculate the same number of diagrams. Since
the total number of diagrams affects the total runtime of each algorithm
and not the ratio of the runtimes, we derive the computational cost
of calculating the signal $S_{d}^{(n)}$ due to a single Feynman diagram
for each algorithm, which we call $C_{\alg}$ and $C_{RKE}$. This
cost is the cost of calling $O_{j^{(*)}}$ $n$ times to arrive at
$\kket{\rho_{d}^{(n)}(t)}$, plus the additional cost of calculating
the signal $S_{d}^{(n)}$ from $\kket{\rho_{d}^{(n)}(t)}$, as in,
for example, Eq.~\ref{eq:S^n}. If the RWA holds, and there are well-defined
manifolds (ground-state, singly excited, doubly excited, etc.) such
that only the optical perturbations couple between them on the timescales
of interest, then the density matrix and Liouvillian can be broken
into smaller pieces. We derive the cost of the general case where
the manifolds are not separable and then extend the result to separable
manifolds.

\subsection{$\protect\alg$\label{subsec:UF2-cost}}

For $\alg$, each evaluation of the $O_{j^{(*)}}$ operators is dominated
by two operations: (1) multiplying the old state by the dipole operator
to obtain $y_{\beta}(t)$ (see Eq.~\ref{eq:Obar-general}), and (2)
performing the convolution $\theta(t)*y_{\beta}(t)$ using the FFT
(see Eq.~\ref{eq:Obar-compact}). The cost of both of these operations
depends on the model system being studied. In the general case with
inter-manifold relaxation processes, all density matrices are expressed
as vectors of length $N^{2}$: 
\[
\kket{\rho}=\sum_{\alpha}c_{\alpha}(t)e^{z_{\alpha}t}\kket{\alpha},
\]
and all the operators on this space, $\mathscr{L}_{0}$, $\boldsymbol{\mu}^{K}\cdot\cvec e_{i}$,
$\boldsymbol{\mu}^{B}\cdot\cvec e_{i}$, are $N^{2}\times N^{2}$
matrices. Given $\kket{\rho^{(n-1)}(t)}$, the first step in determining
$\text{\ensuremath{\kket{\rho_{d}^{(n)}(t)}}=\ensuremath{O_{j^{(*)}}\kket{\rho^{(n-1)}(t)}}}$
is to determine the coefficients $y_{\beta}(t)$ at $M$ time points,
where $M$ must be large enough to well-represent the pulse envelope
shape $A(t)$, see Fig.~\ref{fig:Smallwood-comparison}. This cost
is the cost of the matrix-vector multiplication $\mu^{O}\kket{\rho}$
performed $M$ times, 
\[
C(\mu^{O}\kket{\rho(t)})\sim MC_{\mu}^{\alg},
\]
where we show that for all of the cases we study here, $C_{\mu}^{\alg}$
scales as either $N^{3}$ or $N^{4}$. The next step, calculating
the convolutions $\theta(t)*y_{\beta}(t)$ using the FFT, is 
\[
C\left(\theta(t)*y_{\beta}(t)\right)\sim N^{2}M\log_{2}M,
\]
where the factor of $N^{2}$ arises from the $N^{2}$ values of $\beta$.
Since $C\left(\theta(t)*y_{\beta}(t)\right)\sim N^{2}$, this cost
is lower order than $C_{\mu}$, and we disregard it. Asymptotically
we thus find that
\begin{align}
C(O_{j^{(*)}}) & \sim MC_{\mu}^{\alg},\label{eq:COj_UF2}
\end{align}
where we retain the scaling with $M$ for the purpose of comparison
to RKE later. 

For an $n^{\text{th}}$-order spectroscopy, each Feynman diagram describes
an $n^{\text{th}}$-order density matrix. Starting from the unperturbed
density matrix $\kket{\rho^{(0)}}$, we require $n$ calls to the
$O_{j^{(*)}}$ operator, and so the cost of obtaining $\kket{\rho_{d}^{(n)}(t)}$
from $\kket{\rho^{(0)}}$ is $nC(O_{j^{(*)}})$. This cost accrues
for a single set of pulse delays. Since the calculations of $\kket{\rho^{(n-1)}(t)}$
can be reused, the most expensive part of calculating a multidimensional
spectrum is varying the last pulse delay. We discussed this scaling
in Appendix A of Ref.~\onlinecite{Rose2019}, and the same arguments
apply here.

In order to calculate the desired signal $S_{d}$ from $\kket{\rho_{d}^{(n)}(t)}$,
the density matrix must be evaluated at a single time point (in the
case of integrated measurements, as in phase-cycling experiments)
or at the $M+m_{t}$ time points that determine $P_{d}^{(n)}(t)$
(as in Eq.~\ref{eq:S^n} for phase-matching experiments). $M$ time
points are needed to resolve the turn-on of the signal, governed by
the pulse envelope shape $A(t)$, and $m_{t}$ is determined by the
optical dephasing rate(s) and the desired frequency resolution of
the final signal. Since $P_{d}^{(n)}(t)=\Tr[\boldsymbol{\mu}\rho_{d}^{(n)}(t)]$,
we define the cost $C\left(\Tr[\boldsymbol{\mu}\rho]\right)=C_{\langle\mu\rangle}^{\alg}$
at a single time. Taking the trace at $M+m_{t}$ points has a cost
of $C_{\langle\mu\rangle}^{\alg}(M+m_{t})$. For polarization-based
signals, 
\begin{align}
C(P_{d}^{(n)}(t)) & \sim\underbrace{MC_{\mu}^{\alg}}_{C(O_{j^{(*)}})}+\underbrace{(M+m_{t})C_{\langle\mu\rangle}^{\alg}}_{C\left(\Tr[\boldsymbol{\mu}\rho_{f}^{(n)}(t)]\right)}.\label{eq:CP^n}
\end{align}
The cost of taking the FFT of $P_{d}^{(n)}(t)$ to obtain $S_{d}^{(n)}(\omega)$
does not depend on $N$ and is negligible. Thus 
\begin{equation}
C_{\alg}\sim MC_{\mu}^{\alg}+(M+m_{t})C_{\langle\mu\rangle}^{\alg}\label{eq:UF2-cost}
\end{equation}
per set of pulse delay times. Depending upon the structure of $\mathscr{L}_{0}$,
it is possible that $C_{\mu}^{\alg}=C_{\langle\mu\rangle}^{\alg}$,
in which case calculating the polarization from $\kket{\rho_{d}^{(n)}(t)}$
can be more expensive than constructing $\kket{\rho_{d}^{(n)}(t)}$,
while in other cases, $C_{\langle\mu\rangle}^{\alg}$ may be negligible.
For phase-cycling cases, $C_{\alg}$ is only $MC_{\mu}^{\alg}$.

In order to use the $\alg$ algorithm, we must also diagonalize $\mathscr{L}_{0}$.
We call the cost of this operation $C_{D}$. The scaling of this cost
with $N$ is important for understanding for which Hamiltonian sizes
$\alg$ has an advantage over RKE. However, the precise size $N$
at which the diagonalization cost becomes important depends upon how
many calculations are done using the diagonalization, since $C_{D}$
is amortized over each diagram, each time delay, each electric field
shape studied, and each molecular-frame electric field polarization
considered, as described in Sec.~\ref{sec:Computational-advantage}.

\subsection{RKE \label{subsec:RKE-cost}}

For RKE, each call to $O_{j^{(*)}}$ (1) uses the Euler method to
connect $\rho^{(n-1)}$ to $\rho^{(n)}$ via Eq.~\ref{eq:Euler}
while the pulse is non-zero, and (2) extends the density matrix beyond
$t_{j,\text{max}}$ using the RK45 method to solve the ODE given by
Eq.~\ref{eq:RK45ode}. 

Given any state $\kket{\rho}$, the cost of the evolution according
to Eq.~\ref{eq:EOM} when $\mathscr{L}'(t)=0$ is the cost of multiplying
the vector $\kket{\rho}$ by the matrix $\mathscr{L}_{0}$. Given
a local tolerance $\epsilon$, the RK45 algorithm takes adaptive steps
of size $dt_{RK}$. We approximate $dt_{RK}$ as a constant and neglect
the additional cost incurred when a step is rejected. For each step
$dt_{RK}$, the RK45 algorithm must evaluate $\mathscr{L}_{0}\kket{\rho}$
6 times. We take the cost per time step to be $C_{RK}=6C(\mathscr{L}_{0}\kket{\rho})$. 

Given $\kket{\rho^{(n-1)}(t)}$, the cost of determining $\kket{\rho^{(n)}(t)}$
from $t_{j,\text{min}}$ to $t_{j,\text{max}}$ is the cost of the
two main ingredients: including the pulse via the operation $\mu^{O}\kket{\rho^{(n-1)}}$,
which has cost $C_{\mu}^{RKE}$, and a call to the RK45 algorithm
with cost $C_{RK}$. We divide the interval $[t_{j,\text{min}},t_{j,\text{max}}]$
into $M_{E}$ time points with equal spacing $dt_{E}$. Thus the cost
of determining $\kket{\rho^{(n)}(t)}$ from $t_{j,\text{min}}$ to
$t_{j,\text{max}}$ is 
\[
C_{\text{pulse}}=\left(C_{\mu}^{RKE}+C_{RK}\right)M_{E},
\]
where we have assumed that $dt_{E}<dt_{RK}$. 

From $t_{j,\text{max}}$ to some final time $t_{f}$, the RK45 algorithm
advances the density matrix forward in time with cost $C_{RK}M_{RK}$,
where $M_{RK}=(t_{f}-t_{j,\text{max}})/dt_{RK}$. Thus
\begin{align*}
C_{RKE}(O_{j^{(*)}}) & =C_{\text{pulse}}+C_{RK}M_{RK}\\
C_{RKE}(O_{j^{(*)}}) & =M_{E}C_{\mu}^{RKE}+\left(M_{E}+M_{RK}\right)C_{RK}.
\end{align*}
As with $\alg$, RKE must also resolve the polarization field, which
involves the cost $C_{\langle\mu\rangle}$. However, for RKE, $C_{\langle\mu\rangle}$
is always a lower-order cost. In diabatic Lindblad, the cost of $\Tr[\mu\rho]$
scales linearly with $N$, because $\mu$ is sparse. For both full
and secular Redfield, the cost of $\Tr[\mu\rho]$ scales as $N^{2}$.
In all cases these are lower order than other scaling costs (as summarized
in Table~\ref{tab:Computational-cost-summary} and derived below).
The cost of a signal for RKE is thus
\[
C_{RKE}\sim M_{E}C_{\mu}^{RKE}+\left(M_{E}+M_{RK}\right)C_{RK}
\]

\subsection{Open systems models}

\subsubsection{Diabatic Lindblad}

In the diabatic damping approximation, $\mathscr{L}_{0}$ and $\mu^{O}$
are represented in the site and vibration number basis. In this basis,
$\mathscr{L}_{0}$ and $\mu^{O}$ are sparse matrices, so that for
RKE, $C_{\mu}^{RKE}$ and $C_{RK}$ both scale as $N^{2}$. For $\alg$,
the cost of diagonalization is $C_{D}\sim N^{6}$. Transforming $\mu^{O}$
into the eigenbasis of $\mathscr{L}_{0}$ causes $\mu^{O}$ to become
dense, so that for $\alg$, both $C_{\mu}^{\alg}$ and $C_{\langle\mu\rangle}^{\alg}$
scale as $N^{4}$. Therefore, even without $C_{D}$, RKE outperforms
$\alg$ for large enough $N$. We find that RKE starts to outperform
$\alg$ around $N\approx100$ in our test cases shown in Fig.~\ref{fig:Timing-ratios}.

\subsubsection{Full Redfield}

In the full Redfield formalism, $\mathscr{L}_{0}$ is expressed in
the eigenbasis of $H_{0}$, and is a dense $N^{2}\times N^{2}$ matrix,
which requires both RKE and $\alg$ to work in this eigenbasis; RKE
then no longer has the advantage of a sparse $\mu$ operator. We define
the eigenstates of $H_{0}$ to be $H_{0}\ket i=\epsilon_{i}\ket i$.
In $\mathbb{H}$, when $\mu$ is transformed into the eigenbasis of
$H_{0}$ it becomes a dense $N\times N$ matrix, and thus $\mu^{O}$
in $\mathbb{L}$ is a sparse matrix with $N$ entries per row ($\mu^{O}\kket{\rho}$
can be reexpressed as $\mu\rho$ or $\rho\mu$, which shows more transparently
that this operation is the cost of multiplying two dense matrices,
with cost scaling as $N^{3}$).

Therefore all of the dipole-multiplications have the same scaling,
$C_{RK},C_{\mu}^{\alg},C_{\langle\mu\rangle}^{\alg}\sim N^{4}$ and
$C_{\mu}^{RKE}\sim N^{3}$. RKE is then dominated by the cost of propagating
the density matrix using the RK45 method, and both $C_{RKE}$ and
$C_{\alg}$ scale as $N^{4}$. For $\alg$, $C_{D}\sim N^{6}$, as
before.

At large $N$, we find
\[
\frac{C_{RKE}}{C_{\alg}}\approx\frac{\left(M_{E}+M_{RK}\right)C_{RK}}{(2M+m_{t})C_{\mu}^{\alg}}.
\]
$C_{RK}$ is 6 times the cost of matrix-vector multiplication, while
$C_{\mu}^{\alg}$ is the cost of a single matrix-vector multiplication,
and so in this case $C_{RK}\approx6C_{\mu}^{\alg}$. In the cases
that we have tested, with parameters chosen to achieve 1\% agreement
in the resulting spectra, we typically find that $M_{E}\approx20M$
and that $M_{RK}\approx m_{t}\approx M$. With these substitutions,
for large $N$,
\[
\frac{C_{RKE}}{C_{\alg}}\approx40.
\]
Since $\alg$ has better prefactors than RKE, RKE does not outperform
$\alg$ until $C_{D}$ becomes dominant, though memory constraints
(not included in this calculation) likely constrain $N$ before this
crossover is reached. Figure \ref{fig:Redfield-timing-ratios} shows
that for $N\approx100$, $C_{RKE}/C_{\alg}\approx80$, exceeding the
estimate here, even when including the cost of diagonalization in
the $\alg$ cost.

\subsubsection{Secular Redfield}

In secular Redfield formalism, $\mathscr{L}_{0}$ is expressed in
the eigenbasis of $H_{0}$, but nearly all of the entries of this
$N^{2}\times N^{2}$ matrix are zero. The unitary part of $\mathscr{L}_{0}$
is diagonal in this basis, and so the only off-diagonal terms come
from the Redfield tensor $R$. The secular approximation sets all
terms of $R_{ijkl}$ to zero, except those that satisfy the condition
that $|\omega_{ij}-\text{\ensuremath{\omega}}_{kl}|=0$, where $\omega_{ij}=\omega_{i}-\omega_{j}$
specifies the time evolution frequency of the density matrix element
$\rho_{ij}$ due to the unitary part of $\mathscr{L}_{0}$. All populations
$\rho_{ii}$ evolve at $\omega_{ii}=0$, and thus the secular approximation
preserves all of the terms of $R_{iikk}$. All coherence-coherence
and coherence-population terms are zero unless there are degeneracies
in $H_{0}$ or harmonic ladders of eigenstates \cite{May2011}. Even
in those cases, the subsets of coupled coherences form additional
blocks in $\mathscr{L}_{0}$ that are of a negligible size compared
to the $N\times N$ populations block (see note below). Thus, in the
secular approximation, we have $C_{D}\sim N^{3}$, regardless of the
structure of $H_{0}$.

In the general case without degenerate eigenstates and harmonic ladders,
$\mathscr{L}_{0}$ has a single $N\times N$ block coupling populations
and is otherwise already diagonal. Since $\mathscr{L}_{0}$ is block
diagonal, it is also sparse, and therefore $C_{RK}\sim N^{2}$. As
for full Redfield theory, $C_{\mu}^{RKE}\sim N^{3}$ because the dipole
operator must be represented in the eigenbasis of $H_{0}$.

For $\alg$, we diagonalize $\mathscr{L}_{0}$ by finding the right
and left eigenvectors, $\kket{\alpha}$ and $\bbra{\bar{\alpha}}$,
respectively. Let $V_{R}$ be the matrix whose columns are $\kket{\alpha}$,
and let $V_{L}$ be the matrix whose rows are the $\bbra{\bar{\alpha}}$.
Just as with $\mathscr{L}_{0}$, $V_{R}$ and $V_{L}$ each have a
dense $N\times N$ block, with the rest of each matrix being an identity
(since $\mathscr{L}_{0}$ was otherwise already diagonal). For $\alg$,
we represent $\rho$ in the eigenbasis of $\mathscr{L}_{0}$ as in
Eq.~\ref{eq:rho-L-eigenbasis}. We can also represent $\rho$ in
the basis that arises naturally from the eigenbasis of $H_{0}$ as
\[
\rho=\sum_{i,j}c_{ij}\kket{ij},
\]
where $\kket{ij}=\ket i\bra j$. $V_{R}$ and $V_{L}$ allow us to
move between these two bases. For general $\mathscr{L}_{0}$, as in
the full Redfield or diabatic Lindblad cases, $V_{R}$ and $V_{L}$
are dense, and so the cost of evaluating $V_{L}\kket{\rho}$ or $V_{R}\kket{\rho}$
scales as $N^{4}$; in those cases, we do not move between bases in
order to compute $\mu^{O}\kket{\rho}$, but instead transform the
dipole operator into the $\kket{\alpha}$ basis once. However, in
the secular approximation, the cost of $V_{L}\kket{\rho}$ and $V_{R}\kket{\rho}$
scales as $N^{2}$ and is thus a negligible asymptotic cost. $\alg$
can then propagate in the $\kket{\alpha}$ basis and apply $\mu$
in the $\kket{ij}$ basis, giving a large performance improvement.
Then $C_{\mu}^{\alg}$ is identical to $C_{\mu}^{RKE}$, scaling as
$N^{3}$. Note that in the $\kket{ij}$ basis, $C_{\langle\mu\rangle}^{\alg}\sim N^{2}$,
so is negligible. We then find 
\[
\frac{C_{RKE}}{C_{\alg}}\approx\frac{C_{\mu}^{RKE}M_{E}}{C_{\mu}^{\alg}M}\approx20,
\]
where we have once again used $M_{E}\approx20M$ from the studies
in Fig.~\ref{fig:Redfield-timing-ratios}. This result shows that
$\alg$ always outperforms RKE, regardless of $N$. Thanks to the
block-diagonal structure of $\mathscr{L}_{0}$, $C_{D}$ is unimportant,
regardless of $N$.

\subsubsection{Secular Redfield with harmonic modes}

We now briefly justify the claim that $C_{D}\sim N^{3}$ even for
the case of harmonic ladders. Let us take a system of $k$ harmonic
modes with unique frequencies $\omega_{\alpha}$. Representing each
mode in the number basis, we truncate each mode at an occupation number
of $n$. In this case $N=n^{k}$. In the secular approximation, all
of the populations are coupled, and so, as stated above, $\mathscr{L}_{0}$
has a dense $N\times N$ block describing population dynamics. The
secular approximation only couples the coherences of a single harmonic
mode to other coherences of the same mode. Furthermore, it only couples
coherences that oscillate at the same frequency. In the stated truncation
scheme, each mode has $n-1$ coherences that oscillate at frequency
$\omega_{ij}=\omega_{\alpha}$. Each mode has $n-2$ coherences that
oscillate at frequency $\omega_{ij}=2\omega_{\alpha}$. In general,
each mode has $n-\nu$ coherences that oscillate at frequency $\omega_{ij}=\nu\omega_{\alpha}$.
Therefore, for each harmonic mode, there are $\nu$ blocks of size
$(n-\nu)\times(n-\nu)$ for $\nu=1,2,...n$. The cost of diagonalizing
each block scales as $\sim(n-\nu)^{3}$, and the total cost of diagonalizing
all of the blocks is therefore 
\[
k\sum_{\nu=1}^{\nu}(n-\nu)^{3}\approx kn^{4}.
\]
Now recall that the cost of diagonalizing the population block is
$O(N^{3})$, which is $O(n^{3k})$. Thus for $k>1$, the cost of diagonalizing
all of the smaller coherence-coupling blocks is negligible for computational
complexity analyses. For the case of $k=1$, an analytical solution
exists for diagonalizing $\mathscr{L}_{0}$ and thus $C_{D}=0$ \cite{Rose2012}.
The derivation of the analytical solution is in the Lindblad formalism;
however, secular Redfield can be mapped onto the Lindblad formalism.
Therefore, even for the case of harmonic ladders, we have that $C_{D}\sim N^{3}$.

\subsection{Separable manifolds}

We briefly describe how both $\alg$ and RKE scale when there is no
inter-manifold relaxation process, and therefore $\mathscr{L}_{0}$
breaks down into blocks of size $N_{X}N_{Y}\times N_{X}N_{Y}$, and
thus for dense $\mathscr{L}_{0}$, $C_{RK45}\sim N_{X}^{2}N_{Y}^{2}$.
When $X=Y$, the block describes population and coherence dynamics
within a manifold. When $X\neq Y$, the block describes the evolution
of coherences between manifolds. In general, the dipole operator $\mu^{O}$
connects blocks of $\mathscr{L}_{0}$ by changing either $X$ or $Y$,
and thus has a shape $N_{X}N_{Y'}\times N_{X}N_{Y}$ or $N_{X'}N_{Y}\times N_{X}N_{Y}$.
For dense $\mu^{O}$, $C_{\mu}\sim N_{X}^{2}N_{Y}N_{Y'}$. The actual
scalings depend upon the sparsity structure (or lack thereof) of $\mathscr{L}_{0}$
and $\mu^{O}$. The asymptotic costs in terms of manifold sizes $N_{X}$
are summarized in Table~\ref{tab:Manifold-computational-cost}. We
plot the scaling of $C_{RK}$ and $C_{\alg}$ and their ratios as
a function of $N_{SEM}$ because the ratio $C_{RK}/C_{\alg}$ (1)
depends upon $N_{SEM}$ alone for diabatic Lindblad, (2) depends upon
$N_{DEM}/N_{SEM}$ for full Redfield, or (3) is a constant for secular
Redfield. The runtime costs of $C_{\alg}$ and $C_{RKE}$ individually
depend upon both $N_{SEM}$ and $N_{DEM}$, as shown below.

\begin{table*}
\caption{\label{tab:Manifold-computational-cost}Summary of asymptotic computational
cost for each algorithm and bath formalism. This table has the same
form as Table~\ref{tab:Computational-cost-summary}, with $N$ replaced
with $N_{X}$ and $N_{Y}$, where $X\protect\neq Y$. The cost of
a general $n^{th}$-order spectroscopy scales with the size of the
two largest accessible manifolds. For $3^{rd}$-order spectroscopies
of polymers with $s>2$, the two largest accessible manifolds are
$X=SEM,DEM$.}
\begin{tabular}{cccccc}
 &  &  &  &  & \tabularnewline
\hline 
\hline 
 & $\alg$ Scaling & \multicolumn{2}{c}{RKE Scaling} & $\mathscr{L}_{0}$ Diagonalization Cost & $\alg$ advantage\tabularnewline
\hline 
 &  & $C_{RK45}$ & $C_{\mu}^{RKE}$ &  & \tabularnewline
Full Redfield & $N_{X}^{2}N_{Y}N_{Y'}$ & $N_{X}^{2}N_{Y}^{2}$ & $N_{X}^{2}N_{Y}$ & $N_{X}^{3}N_{Y}^{3}$ & Up to large $N_{X}$\tabularnewline
Secular Redfield & $N_{X}^{2}N_{Y}$ & $N_{X}N_{Y}$ & $N_{X}^{2}N_{Y}$ & $N_{X}^{3}$ & All $N_{X}$\tabularnewline
Diabatic Lindblad  & $N_{X}^{2}N_{Y}N_{Y'}$ & $N_{X}N_{Y}$ & $N_{X}N_{Y}$ & $N_{X}^{3}N_{Y}^{3}$ & $N_{X}\lesssim100$\tabularnewline
\hline 
 &  &  &  &  & \tabularnewline
\end{tabular}
\end{table*}

\subsubsection{Full Redfield}

In full Redfield theory, RKE is dominated by the RK45 algorithm, with
$C_{RK45}\sim N_{X}^{2}N_{Y}^{2}$. For $3^{rd}$-order spectroscopies,
the most expensive diagram is the ESA, which evolves $\rho^{(3)}$
in the coherence between the $X=SEM$ and the $Y=DEM$, so $C_{RK45}\sim N_{SEM}^{2}N_{DEM}^{2}$.

$\alg$ is dominated by $C_{\mu}^{\alg}\sim N_{X}^{2}N_{Y}N_{Y'}$.
For the ESA, $\alg$ is dominated by the cost of $\mu^{K}\kket{\rho^{(2)}}$,
where $\rho^{(2)}$ is in the SEM with length $N_{SEM}^{2}$. $\mu^{K}$
connects the SEM block to the SEM/DEM coherence block, and so has
shape $N_{SEM}N_{DEM}\times N_{SEM}^{2}$. Thus $C_{\mu}^{\alg}\sim N_{SEM}^{3}N_{DEM}$.

\subsubsection{Secular Redfield}

Both $\alg$ and RKE are dominated by the same operation in the asymptotic
limit, evaluating $\mu\rho^{(n-1)}$. In secular Redfield both $\alg$
and RKE perform this multiplication with $\mu$ and $\rho$ represented
as matrices in the eigenbasis of $H_{0}$. In general $\mu$ has size
$N_{X'}\times N_{Y}$ and $\rho^{(n-1)}$ has size $N_{X}\times N_{Y}$.
For the ESA, $\mu$ has size $N_{DEM}\times N_{SEM}$, and $\rho^{(2)}$
has size $N_{SEM}\times N_{SEM}$. The cost of this operation then
scales in general as $C_{\mu}^{\alg}\sim C_{\mu}^{RKE}\sim N_{X}^{2}N_{Y}$
and for the ESA as $\sim N_{SEM}^{2}N_{DEM}$.

\subsubsection{Diabatic Lindblad}

In diabatic Lindblad, $\mathscr{L}_{0}$ is sparse. RKE represents
$\mu^{O}$ in the diabatic site and vibration-number basis, so that
it is also sparse. Thus $C_{RK45}\sim C_{\mu}^{RKE}\sim N_{X}N_{Y}$.
The cost of the ESA then goes as $\sim N_{SEM}N_{DEM}$.

In the eigenbasis of $\mathscr{L}_{0}$, $\mu^{O}$ is dense, and
so, just as for full Redfield theory, $C_{\mu}^{\alg}\sim N_{X}^{2}N_{Y}N_{Y'}$,
and for the ESA, $C_{\mu}^{\alg}\sim N_{SEM}^{3}N_{DEM}$.

\bibliographystyle{aipnum4-1}
\bibliography{OpenUF2Paper}

\end{document}